\documentclass[%
 aip,
 amsmath,amssymb,
 reprint,%
]{revtex4-1}

\usepackage{graphicx}
\usepackage{dcolumn}
\usepackage{bm}

\usepackage[utf8]{inputenc}
\usepackage[T1]{fontenc}
\usepackage{mathptmx}
\usepackage{booktabs}
\usepackage{etoolbox}
\usepackage{lipsum} 
\newcommand{\p}{\partial}
\usepackage{xcolor}
\usepackage{tikz}

\usepackage[justification=justified,format=plain]{caption}
\makeatletter
\def\@email#1#2{%
 \endgroup
 \patchcmd{\titleblock@produce}
  {\frontmatter@RRAPformat}
  {\frontmatter@RRAPformat{\produce@RRAP{*#1\href{mailto:#2}{#2}}}\frontmatter@RRAPformat}
  {}{}
}%
\usepackage{subcaption}
\usepackage{overpic}
\makeatother
\begin{document}

\preprint{AIP/123-QED}

\title[Peristalsis in perivascular networks]{The directional flow generated by peristalsis in perivascular networks -- theoretical and numerical reduced-order descriptions}

\author{I. Gjerde}%
\email{ingeborg@simula.no}
\affiliation{Department of Numerical Analysis and Scientific Computing, Simula Research Laboratory, Kristian Augusts gate 23, 0164 Oslo, Norway}

\author{M. E. Rognes}%
\email{meg@simula.no}
\homepage{https://www.simula.no/people/meg}
\affiliation{Department of Numerical Analysis and Scientific Computing, Simula Research Laboratory, Kristian Augusts gate 23, 0164 Oslo, Norway}

\author{A.L. S\'anchez}
\email{als@ucsd.edu.}
\homepage{http://www.http://asanchez.ucsd.edu/}
\affiliation{Department of Mechanical and Aerospace Engineering, University of California San Diego, La Jolla, US 
}

\date{\today}

\begin{abstract}
Directional fluid flow in perivascular spaces surrounding cerebral arteries is hypothesized to play a key role in brain solute transport and clearance. While various drivers for pulsatile flow, such as cardiac or respiratory pulsations, are well quantified, the question remains as to which mechanisms could induce directional flow within physiological regimes. To address this question, we develop theoretical and numerical reduced-order models to quantify the directional (net) flow induceable by peristaltic pumping in periarterial networks. Each periarterial element is modeled as a slender annular space bounded internally by a circular tube supporting a periodic traveling (peristaltic) wave. Under the reasonable assumptions of small Reynolds number flow, small radii, and small-amplitude peristaltic waves, we use lubrication theory and regular perturbation methods to derive theoretical expressions for the directional net flow and pressure distribution in the perivascular network. The reduced model is used to derive closed-form analytical expressions for the net flow for simple network configurations of interest, including single elements, two elements in tandem, and a three element bifurcation, with results compared with numerical predictions. In particular, we provide a computable theoretical estimate of the net flow induced by peristaltic motion in perivascular networks as a function of physiological parameters, notably wave length, frequency, amplitude and perivascular dimensions. Quantifying the maximal net flow for specific physiological regimes, we find that vasomotion may induce net pial periarterial flow velocities on the order of a few to tens of $\mu$m/s and that sleep-related changes in vasomotion pulsatility may drive a threefold flow increase.
\end{abstract}

\maketitle

\section{Introduction}
\label{sec:introduction}

The pulsatile motion of cerebrospinal fluid (CSF) in the perivascular spaces (PVSs) surrounding cerebral arteries is a complex multiscale phenomenon that has been reasoned to play an important role in brain solute transport and clearance~\cite{flexner1933some, rennels1985evidence, ichimura1991distribution, iliff2012paravascular, iliff2013cerebral}. Understanding and potentially modulating molecular transport in and around the brain is fundamental in the context of brain cancer~\cite{ngo2022perivascular, lilius2023glymphatic}, neurodegenerative diseases~\cite{nedergaard2020glymphatic} such as Alzheimer's disease~\cite{weller2008perivascular, mestre2022periarteriolar} or Parkinson's disease~\cite{zhang2023interaction}, as well as in stroke and other neurological disorders~\cite{rasmussen2018glymphatic}. Intriguingly, molecular transport in the brain is altered by lifestyle factors such as exercise~\cite{von2018voluntary} or sleep~\cite{xie2013sleep, bojarskaite2023sleep}. In spite of its importance, perivascular flow and transport remains enigmatic and only partially quantified. 

The PVSs are spaces or potential spaces that run along blood vessels on the brain surface and within the brain parenchyma, filled with CSF or interstitial fluid (ISF). Their shapes, sizes and hydraulic properties such as permeability or resistance remain under debate~\cite{zhang1990interrelationships, bedussi2018paravascular, wardlaw2020perivascular, tithof2019hydraulic}, with recent evidence indicating substantial variability in their characteristics~\cite{raicevic2023sizes, mestre2022periarteriolar}. They are often represented via annular or elliptic cross-sections surrounding the blood vessels, isolated from or extending into the surrounding subarachnoid space (in the case of pial PVSs)~\cite{bilston2003arterial, tithof2019hydraulic, vinje2021brain, carr2021peristaltic}, or bounded by astrocyte endfeet (in the case of parenchymal PVSs). Perivascular CSF flow is well-approximated by the flow of an incompressible Newtonian fluid at low Reynolds numbers, with peak speeds estimated at up to 40 $\mu$m/s~\cite{mestre2018flow}.  

Biophysics-based modelling of perivascular pathways has seen a surge of interest over the last decade in particular. Mathematical and computational models now provide new insights into the mechanisms underlying perivascular flow and transport complementing experimental and clinical studies~\cite{daversin2020mechanisms, kelley2022glymphatic, bojarskaite2023sleep, vinje2023human}. A key question is how and to what extent physiological pulsations induce oscillatory and directional fluid flow in the perivascular spaces. While arterial pulsations are clearly implicated in driving perivascular flow~\cite{mestre2018flow}, the frequencies and length scales involved have argued against peristaltic pumping associated with the cardiac cycle as an effective mechanism for \emph{directional} (net) fluid flow and transport~\cite{asgari2016glymphatic, martinac2020computational, daversin2020mechanisms, kedarasetti2020arterial}. 

However, the perivascular environment pulsates in synchrony with several different physiological rhythms spanning different scales in space and time. At a frequency of around 1 Hz in humans at rest and up to 10 Hz in mice, the cardiac pulse wave travels along the vascular tree at a wave speed of around 1 m/s, and with changes in the vascular diameter of 1-2\%~\cite{mestre2018flow}. On the other hand, vasomotion, defined as a spontaneous or stimulus-evoked change in vascular diameter at frequencies around 0.1 Hz, have been observed as propagating along pial arterioles of mice at wave speeds around 400 $\mu$m/s and diameter changes on the order of 5-15\%~\cite{van2020vasomotion, munting2023spontaneous}. Third, sleep is associated with changes in perivascular transport~\cite{xie2013sleep, ma2019rapid}, changes in perivascular dynamics with slow large-amplitude oscillations in non-REM sleep and vasodilations during REM sleep~\cite{bojarskaite2023sleep}, as well as brain-wide vasomotor and respiratory pulsations during non-REM sleep~\cite{helakari2022human}. A central question to be addressed below is whether arterial pulsations at these scales in time and space may drive significant net directional flow in a perivascular network.

While the cerebral perivascular network involves multiple branches resulting from subsequent bifurcations (on the order of nine in the mouse brain \cite{blinder2010topological} and more in the human brain), most theoretical or computational descriptions of the associated quasi-steady low-Reynolds number flow focus on individual perivascular elements of uniform cross section extending between two end points with prescribed pressure. By way of contrast, our paper presents a unified theoretical and numerical investigation of the net perivascular flow induced by peristaltic pumping in complex perivascular networks involving multiple branches, leveraging simplifications stemming from the disparity of length scales present in the problem. 

In the reduced-order description, presented in Section~\ref{sec:theory}, each individual periarterial element is modeled as a slender annular space bounded internally by a circular artery supporting a periodic traveling wave. The lubrication limit is used to derive an expression relating the flow rate along a given element and the pressure difference between its ends. For small-amplitude peristaltic waves, the limit of interest in perivascular motion, regular perturbation methods can be used to simplify the solution, which is then used to derive a system of linear equations for the quantification of peristaltic motion in complex networks. The results are used to derive analytical expressions for the flow rate in simple configurations. The theoretical predictions are validated in Section~\ref{sec:numerics} through comparisons with numerical simulations. 

The theoretical results are used in Section~\ref{sec:physiological} to investigate net perivascular flow for parametric values of physiological relevance. Here, the theoretical predictions are also compared with other experimental data. While the cardiac cycle is found to produce negligibly small flow rates, in agreement with previous findings~\cite{asgari2016glymphatic, martinac2020computational, daversin2020mechanisms, kedarasetti2020arterial}, the maximum estimated steady velocities induced by vasomotion in mice are found to be on the order of those observed in previous in-vivo experimental studies~\cite{mestre2018flow}. Concluding remarks are provided in Section~\ref{sec:conclusions} along with a discussion of the limitations and potential extensions of the reduced model.

\section{Derivation of the analytical reduced-order model}
\label{sec:theory}

The pulsatile motion in the cerebral periarterial system is characterized by negligibly small values of the Reynolds number and arterial wavelengths that are large compared with the characteristic transverse dimension, so that the lubrication approximation can be employed to simplify the solution \cite{shapiro1969peristaltic}. In building a network model, it is convenient to begin by analyzing the motion in the perivascular space extending between subsequent arterial bifurcations to determine the relation between the flow rate, the existing inter-bifurcation pressure difference, and the peristaltic wave propagating along the artery. 

\subsection{Problem formulation}
\label{sec:problem}
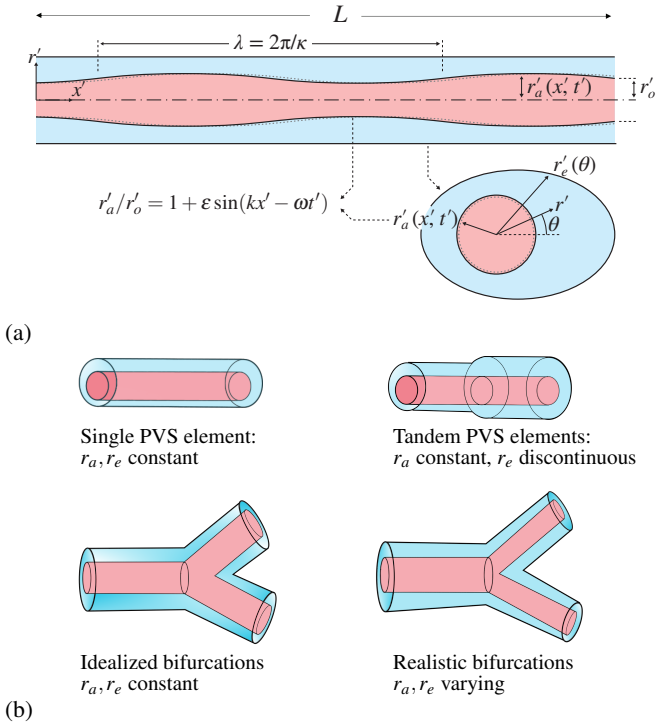
\begin{figure}

\begin{subfigure}{0.5\textwidth}
\includegraphics[width=8.5cm]{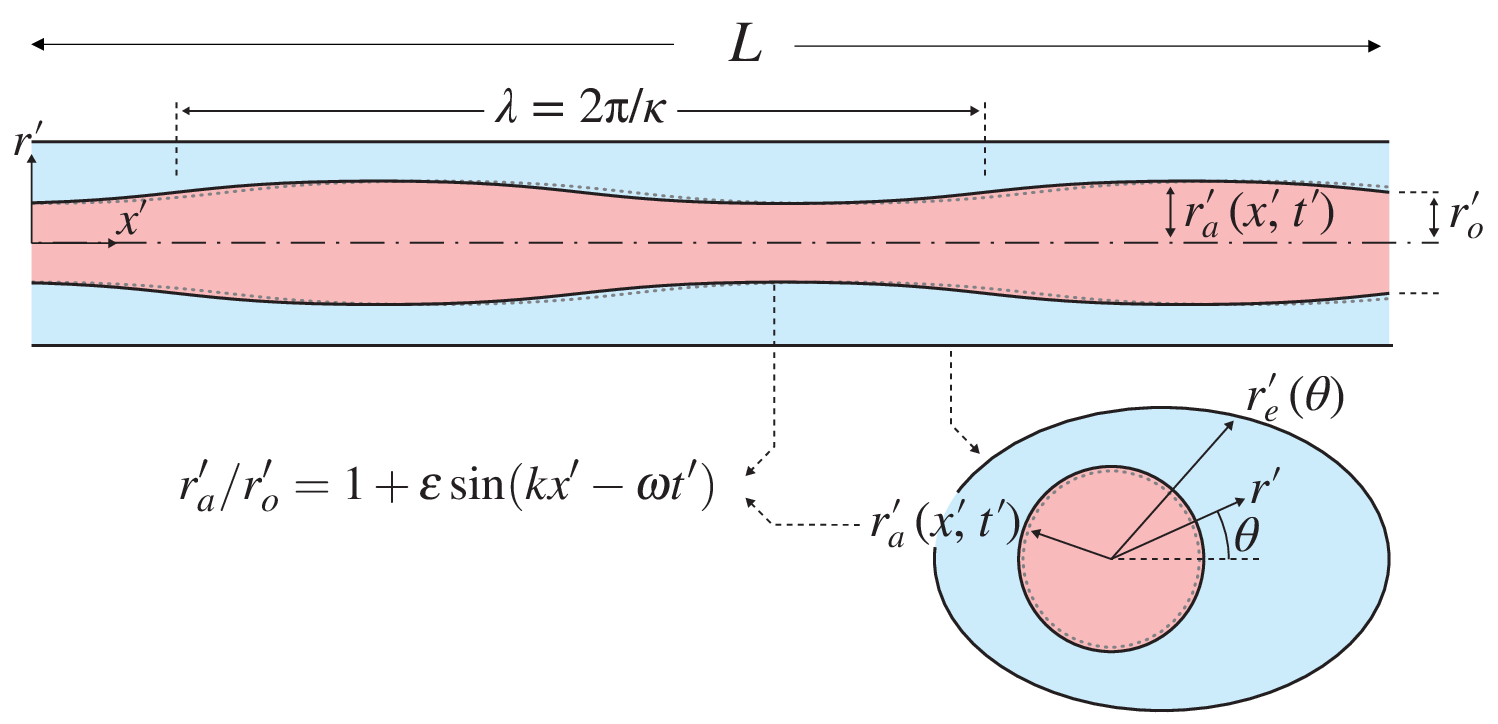}
\caption{} 
\label{fig:1} 
\end{subfigure}
\vspace{1em}
\begin{subfigure}{0.5\textwidth}
\begin{overpic}[width=0.8\textwidth]{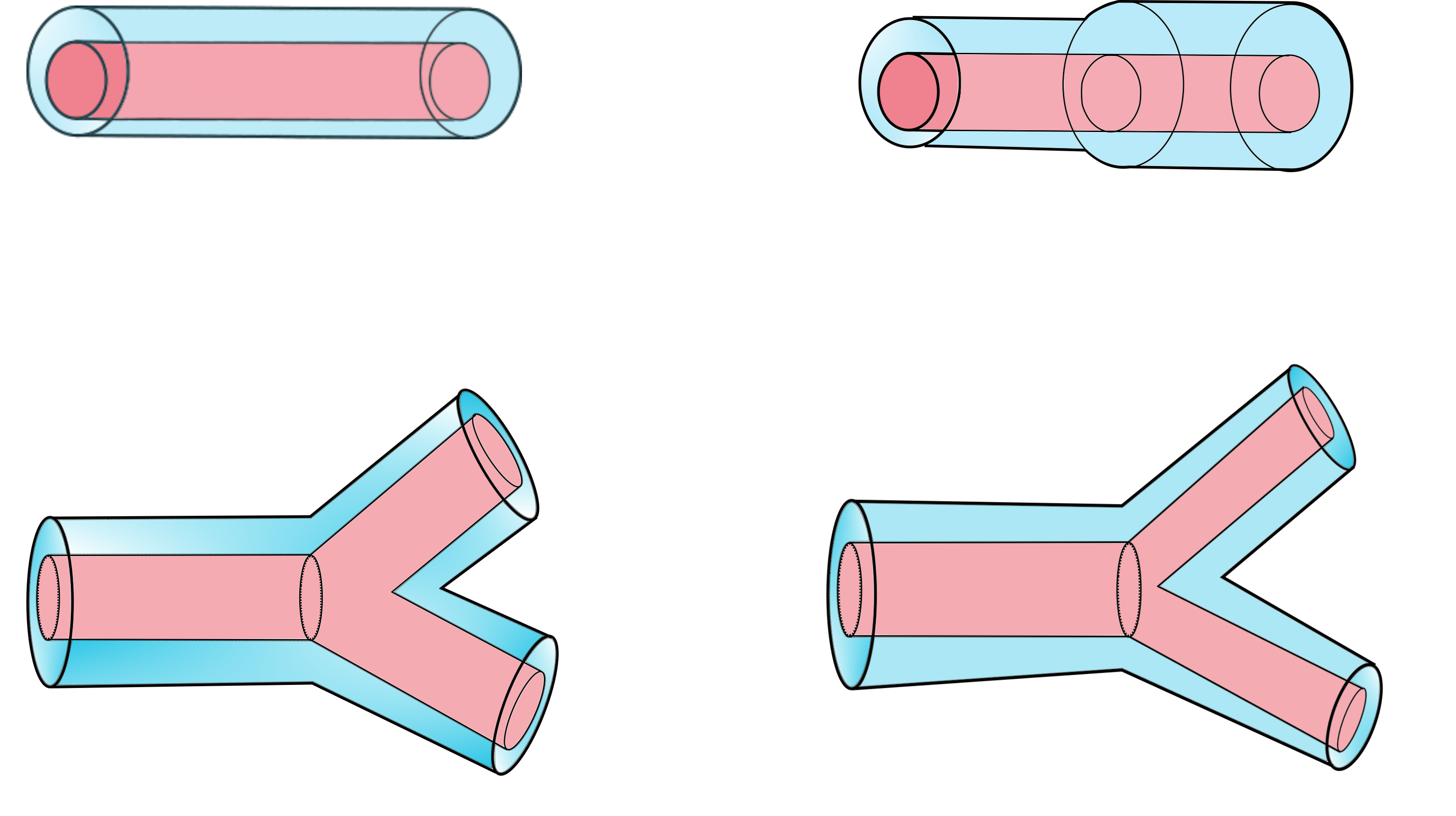}
\put(2,40){\footnotesize Single PVS element:}
\put(2,36){\footnotesize $r_a, r_e$ constant}
\put(60,40){\footnotesize Tandem PVS elements:}
\put(60,36){\footnotesize $r_a$ constant, $r_e$ discontinuous}
\put(2,-2){\footnotesize Idealized bifurcations}
\put(60,-2){\footnotesize Realistic bifurcations}
\put(2,-6){\footnotesize $r_a, r_e$ constant}
\put(60,-6){\footnotesize $r_a, r_e$ varying}
\end{overpic}
\vspace{1em}
\caption{}
\label{fig:1B}
\end{subfigure}
\caption{Peristaltic pumping is modelled by a travelling wave that changes the arterial radius in time and space (Fig.~\ref{fig:1}). The effect of peristaltic pumping will be shown to depend on the network configuration. In this work, we focus on the four cases sketched in Fig.~\ref{fig:1B}.}
\end{figure}

The seminal analysis of peristaltic motion driven by a train of waves travelling along infinite tubes \cite{shapiro1969peristaltic} has been extended to account for multiple effects. Recent work addresses, for example, effects of magnetic fields (see \cite{akram2023hybridized} and references therein). Peristaltic motion in circular tubes of finite length was first investigated by Li and Brasseur \cite{li1993non}. The case of annular tubes was investigated numerically by Carr et al.~\cite{carr2021peristaltic}. The analytic description of the flow is due to Coenen et al. \cite{coenen2021lubrication}, who used the model depicted in Fig.~\ref{fig:1}, including a fixed cylindrical outer boundary with radial distribution $r'_e(\theta)$ and an inner flexible tube of radius
\begin{equation} \label{rakx}
r'_a/r'_o=1+\varepsilon \sin(k x'-\omega t'),
\end{equation}
representing the arterial wall, which supports a traveling wave of relative amplitude $\varepsilon$, wave number $k$ and angular frequency $\omega$. In the following, we shall assume that the wavelength $\lambda=2\pi/k$ is comparable to the element length $L$, so that $\ell=k L =O(1)$. The problem is to be solved for a known value of the pressure difference between the two ends of the tube $p'_L-p'_0$, a function of time. 

The solution is described with use of the cylindrical coordinates $(x',r',\theta)$ indicated in Fig.~\ref{fig:1}, with corresponding velocity components $(u',v',w')$ satisfying the nonslip condition $u'=v'=w'=0$ at $r'=r'_e$ and $u'=v'-\p r'_a/\p t' =w'=0$ at $r'=r'_a$, which can be used in integrating the continuity equation across the perivascular space to give
\begin{equation}
\frac{\p Q'}{\p x'}+2\pi\varepsilon \omega r_o^{'2} [1+\varepsilon \sin(k x'-\omega t')] \cos(k x'-\omega t')=0, \label{cont_int}
\end{equation}
where 
\begin{equation} 
Q'=\int_0^{2\pi}\left(\int_{r'_a}^{r'_e} r' u' \, {\rm d} r'\right) {\rm d}\theta
\end{equation}
is the instantaneous value of the volumetric flow rate.

The boundary condition $v'=\p r'_a/\p t'$ at $r'=r'_a$ can be used together with the expression~\eqref{rakx} to obtain an estimate for the characteristic value $v'_c=\varepsilon r'_o \omega$ of the transverse velocity. It then follows from continuity that the characteristic longitudinal velocity is $u'_c=v'_c \lambda/r'_o=\varepsilon \omega \lambda$. Our analysis addresses peristaltic waves with wavelengths $\lambda$ much larger than the characteristic transverse length $r'_o$, yielding slender flows with $v'_c \ll u'_c$. With the streamlines being nearly aligned with the artery, one can show using the momentum equation that the transverse pressure variations across the perivascular space are a factor $(r'_o/\lambda)^2 \ll 1$ smaller than the corresponding longitudinal variations, so that in the first approximation one can write $p'(x',t')$ for the pressure, independent of $r'$ and $\theta$. Additional simplifications arise in configurations with frequencies $\omega \ll \nu/r_o^{\prime 2}$, where $\nu$ is the fluid kinematic viscosity, such that the characteristic viscous time $r_o^{\prime 2}/\nu$ is much smaller than the characteristic oscillation time $\omega^{-1}$ and also much smaller than the characteristic convective time $\lambda/u'_c=\varepsilon^{-1} \omega^{-1} $. Under those conditions, acceleration is negligibly small, so that the motion is determined by a balance between viscous and pressure forces, with the longitudinal momentum equation reducing to the familiar lubrication form
\begin{equation}
-\frac{1}{\mu} \frac{\p p'}{\p x'}+\frac{1}{r'} \frac{\p}{\p r'}\left(r' \frac{\p u'}{\p r'}\right)+\frac{1}{r^{\prime 2}} \frac{\p^2 u'}{\p \theta^2}=0, \label{mom0}
\end{equation}
with $\mu $ representing the fluid viscosity. Note that the term $\p^2 u'/\p x^{\prime 2}$ has been neglected above when writing the viscous force, as is consistent in the limit $\lambda \gg r'_o$ considered here.

\subsection{General solution}

In describing the solution, periodic in time but non-periodic in $x'$, it is conveniently to introduce the dimensionless independent variables $t=\omega t'$, $\xi=k x'-\omega t'$ and $r=r'/r'_o$, so that the radial boundaries of the perivascular space become $r_a=r'_a/r'_o=1+\varepsilon \sin \xi$ and $r_e(\theta)=r'_e/r'_o$, along with the dimensionless axial velocity $u=u'/u'_c=u'/(2 \pi \varepsilon \omega/k)$ and accompanying flow rate 
\begin{equation} \label{Q_eq}
Q(\xi,t)=\frac{Q'}{2 \pi \varepsilon \omega r_o^{'2}/k}=\int_0^{2\pi}\left(\int_{r_a}^{r_e} r u \, {\rm d} r\right) {\rm d}\theta.
\end{equation}
In terms of these variables, the integrated form of the continuity equation~\eqref{cont_int} becomes 
\begin{equation}
\frac{\p Q}{\p \xi}+(1+\varepsilon \sin \xi) \cos\xi=0, \label{cont_int2}
\end{equation}
while the axial component of the momentum equation~\eqref{mom0} takes the dimensionless form
\begin{equation}
-\frac{\p p}{\p \xi}+\frac{1}{r} \frac{\p}{\p r}\left(r \frac{\p u}{\p r}\right)+\frac{1}{r^2} \frac{\p^2 u}{\p \theta^2}=0, \label{mom}
\end{equation}
where $p(\xi,t)=(p'-p'_0)/[2 \pi\mu \varepsilon \omega/(k r'_o)^2]$ is the pressure variation from the entrance scaled with its characteristic value $\mu \varepsilon \omega/(k r'_o)^2$. Since $p'=p'_0$ at $x'=0$ and $p'=p'_L$ at $x'=L$, it follows that $p=0$ at $\xi=-t$ and $p=\delta p$ at $\xi=\ell-t$, where $\delta p(t)=(p'_L-p'_0)/[2 \pi\mu \varepsilon \omega/(k r'_o)^2]$.

The integration of~\eqref{mom} subject to the nonslip conditions $u=0$ at $r=r_a$ and at $r=r_e$ is facilitated by introduction of the ansatz\cite{white2006viscous} $u=-(\p p/\p \xi) U$, where $U(r,\theta)$ satisfies the Poisson problem
\begin{equation}
\frac{1}{r} \frac{\p}{\p r}\left(r \frac{\p U}{\p r}\right)+\frac{1}{r^2} \frac{\p^2 U}{\p \theta^2}=-1; \quad U=0 \; \; {\rm at} \;\left\{\begin{array}{l} r=r_a \\ r=r_e(\theta) \end{array} \right.. \label{mom2}
\end{equation}
In the solution, the associated flow rate~\eqref{Q_eq} is given by
\begin{equation} \label{Q_R}
Q=-\frac{1}{\mathcal{R}} \frac{\p p}{\p \xi},
\end{equation}
involving the hydraulic resistance
\begin{equation} \label{Rdef}
\mathcal{R}=\left[\int_0^{2\pi}\left(\int_{r_a}^{r_e} r U \, {\rm d} r\right) {\rm d}\theta\right]^{-1},
\end{equation}
a function of $\xi$ through $r_a=1+\varepsilon \sin \xi$.

For an outer boundary of general form $r_e(\theta)$, numerical integration is needed to determine $U(r,\theta;r_a)$ and therefore $\mathcal{R}$. The sample computations presented below consider the particular case $r_e=\alpha \cos\theta+(\beta^2-\alpha^2 \cos^2 \theta)^{1/2}$, corresponding to a circular cylinder of radius $\beta r'_o$ whose center is displaced from the center of the inner cylinder by a distance $\alpha r'_o$, for which an analytical solution is available, with the associated hydraulic resistance described with excellent accuracy by the approximate expression\cite{white2006viscous}
\begin{equation} \label{Rapprox}
\mathcal{R}^{-1}=\frac{\pi}{8} \left[1+\frac{3}{2} \left(\frac{\alpha}{\beta-r_a}\right)^2\right]\left[\beta^4-r_a^4 -\frac{(\beta^2-r_a^2)^2}{\ln(\beta/r_a)}\right]
\end{equation}
with $r_a=1+\varepsilon \sin \xi$.

Substitution of~\eqref{Q_R} into~\eqref{cont_int2} yields the second-order equation
\begin{equation}
\frac{\p}{\p \xi}\left(-\frac{1}{\mathcal{R}} \frac{\p p}{\p \xi}\right)+(1+\varepsilon \sin \xi) \cos\xi=0
\end{equation}
to be integrated with boundary conditions $p=0$ at $\xi=-t$ and $p=\delta p(t)$ at $\xi=\ell-t$. A first integral provides
\begin{equation} \label{QC0}
Q=-\frac{1}{\mathcal{R}} \frac{\p p}{\p \xi}=C(t)-\sin \xi-\frac{\varepsilon}{2} \sin^2 \xi,
\end{equation}
where the function
\begin{equation}\label{C_eq}
C(t)=\frac{\int_{-t}^{\ell-t} \mathcal{R} \sin \xi  \,{\rm d}\xi+(\varepsilon/2) \int_{-t}^{\ell-t} \mathcal{R} \sin^2 \xi  \,{\rm d}\xi - \delta p}{\int_{-t}^{\ell-t} \mathcal{R} {\rm d}\xi}
\end{equation} 
is obtained by integrating a second time to give
\begin{equation}\label{p_eq}
p=-C \int_{-t}^{\xi} \mathcal{R} {\rm d}\xi+\int_{-t}^{\xi} \mathcal{R} [\sin \xi +(\varepsilon/2) \sin^2 \xi] {\rm d}\xi
\end{equation}
and using the known value of the pressure $p=\delta p$ at the tube end $\xi=\ell-t$. It is convenient to rewrite~\eqref{QC0} in terms of the coordinate $x=k x'$ to give
\begin{equation} \label{QC}
Q=C(t)-\sin(x-t)-\frac{\varepsilon}{2} \sin^2(x-t).
\end{equation}
Taking the time average 
\begin{equation}\label{time average}
\langle * \rangle=\frac{\omega}{2\pi} \int_{t'}^{t'+2\pi/\omega}  * \, {\rm d}t' =\frac{1}{2\pi} \int_t^{t+2\pi}  * \, {\rm d}t 
\end{equation}
yields
\begin{equation} \label{pumping}
\langle Q \rangle=\langle C \rangle-\varepsilon/4,
\end{equation}
uniform along the tube.

The above results can be used to evaluate the pumping efficiency of the perivascular space, which can be written following \cite{shapiro1969peristaltic} as
\begin{equation}
E=\frac{\langle Q' (p'_L-p'_0) \rangle}{\left\langle \int_0^L 2 \pi r'_a \frac{\p r'_a}{\p t'} (p'-p'_0) {\rm d} x'\right\rangle},
\end{equation}
where the numerator is the rate at which energy is stored in the fluid and the denominator is the rate at which mechanical work is delivered to the wall, both quantities being averaged over a period. The above expression can be cast in the dimensionless form
\begin{equation}
E= \frac{-\langle Q \, \delta p \rangle}{\left\langle \int_{-t}^{\ell-t} (1+\varepsilon \sin \xi) \cos(\xi) p \, {\rm d} \xi \right\rangle},
\end{equation}
with $p$ and $Q$ given in~\eqref{p_eq} and~\eqref{QC}, respectively.

\subsection{Simplifications for waves of small amplitude}

\begin{figure}
\includegraphics[width=0.4\textwidth]{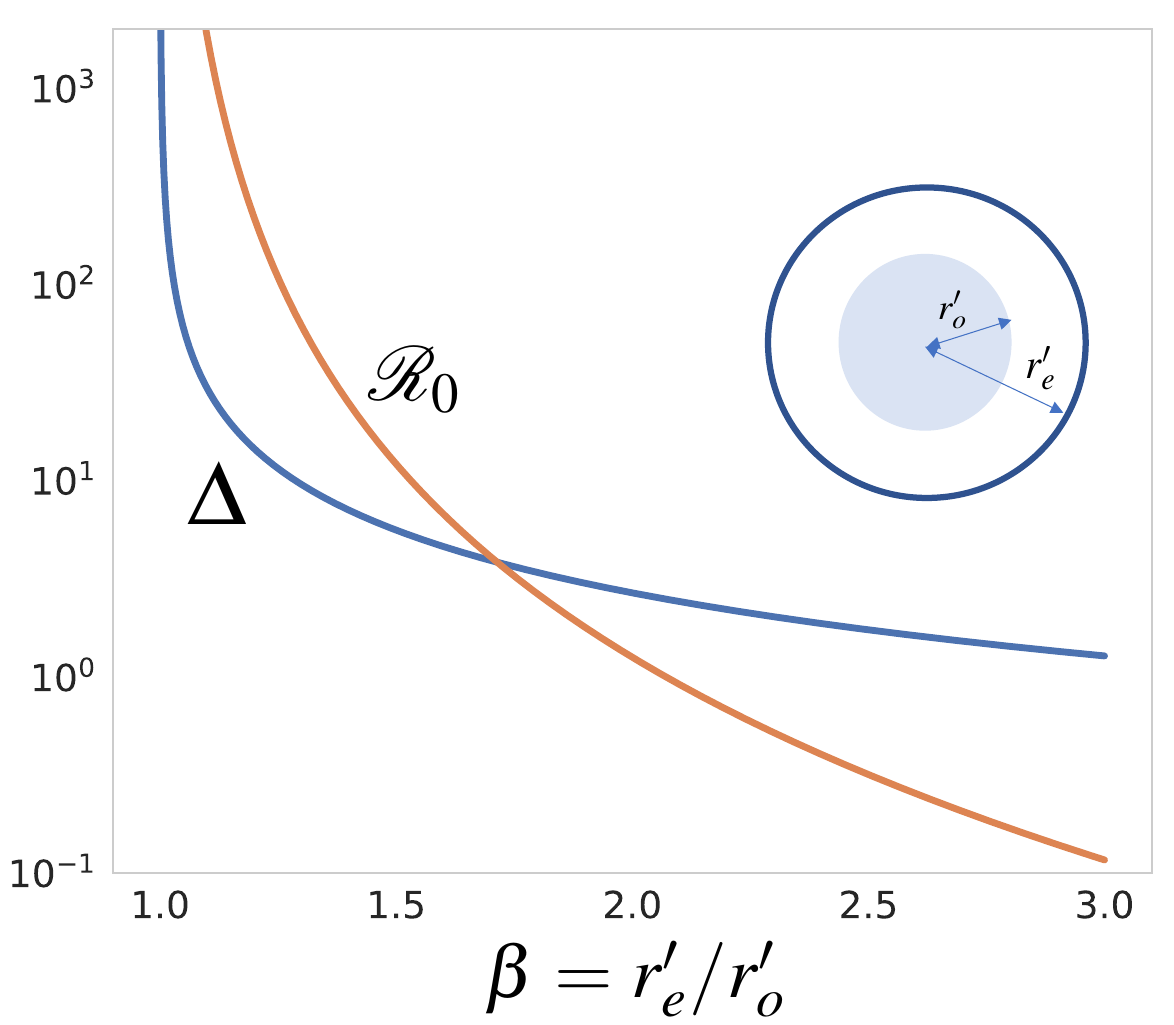}
\caption{The variation of $\mathcal{R}_o$ and $\Delta$ with the outer-to-inner ratio $\beta = r'_e/r'_o$ for a peristaltic element of concentric annular section with inner radius $r'_o$ and outer radius $r'_e$.}
\label{fig:RDelta}
\end{figure}

The integrals in~\eqref{C_eq} can be evaluated numerically with use of~\eqref{Rapprox} and $r_a=1+\varepsilon \sin \xi$. Analytic expressions can be derived in the limit $\varepsilon \ll 1$ by expressing the different quantities as expansions in powers of $\varepsilon$. For instance, the hydraulic resistance~\eqref{Rdef} can be written as 
\begin{equation} \label{Rexp}
\mathcal{R}/\mathcal{R}_0=1+\varepsilon \Delta \sin \xi + O(\varepsilon^2)
\end{equation}
where
\begin{equation}
\mathcal{R}_0=\left. \mathcal{R} \right|_{r_a=1} \quad {\rm and} \quad \Delta=\left.\frac{{\rm d} \mathcal{R}/{\rm d} r_a}{\mathcal{R}} \right|_{r_a=1}
\end{equation}
take the form
\begin{equation}
\mathcal{R}_o^{-1}=\frac{\pi}{8} \left[1+\frac{3}{2} \left(\frac{\alpha}{\beta-1}\right)^2\right] \left[\beta^4-1 -\frac{(\beta^2-1)^2}{\ln(\beta)}\right], 
\end{equation}
and
\begin{equation} \label{Delta_eq}
\Delta=\frac{\left[1+\frac{3}{2} \left(\frac{\alpha}{\beta-1}\right)^2\right]  \left[4+\frac{\beta^2-1}{\ln \beta}\left(\frac{\beta^2-1}{\ln \beta}-4\right)\right]-\frac{3\alpha^2}{(\beta-1)^3}}{\left[1+\frac{3}{2} \left(\frac{\alpha}{\beta-1}\right)^2\right] \left[\beta^4-1 -\frac{(\beta^2-1)^2}{\ln(\beta)}\right]}
\end{equation}
when~\eqref{Rapprox} is used in the evaluation, with the simpler expressions
\begin{equation}
\mathcal{R}_o^{-1}=\frac{\pi}{8} \left[\beta^4-1 -\frac{(\beta^2-1)^2}{\ln(\beta)}\right], 
\end{equation}
and
\begin{equation}
\label{Delta_eq2}
\Delta=\frac{[2-(\beta^2-1)/\ln \beta]^2}{\beta^4-1-(\beta^2-1)^2/\ln \beta}
\end{equation}
applying in the case of concentric cylinders with outer-to-inner radii ratio $\beta=r'_e/r'_o$. As shown in Fig.~\ref{fig:RDelta}, the parameters $\mathcal{R}_o$ and, to a lesser extent, $\Delta$ exhibit a strong dependence on $\beta$, indicating that the shape of the perivascular-space cross section is a critical factor in the associated pumping efficiency.

The expression~\eqref{Rexp} can be used to express the integrals in~\eqref{C_eq} in power expansions of $\varepsilon$. Substituting the result in~\eqref{QC} with $Q=Q_0+\varepsilon Q_1+\cdots$ and $\delta p=\delta p_0+\varepsilon \delta p_1+\cdots$ and collecting terms of order unity yields
\begin{equation} \label{Q_0}
Q_0=-\frac{\delta p_0}{\mathcal{R}_o \ell}+\frac{1}{\ell}[\cos t-\cos(t-\ell)]+\sin(t-x)
\end{equation}
for the leading-order flow rate, whose time-averaged value reduces to $\langle Q_0 \rangle=\langle\delta p_0\rangle/(\mathcal{R}_o \ell)$. On the other hand, the expression for $Q_1$, obtained at the following order, can be time-averaged to give
\begin{eqnarray} \label{Q_1}
\langle Q_1 \rangle&=&-\frac{\langle\delta p_1 \rangle}{\mathcal{R}_o \ell}+\Delta \left(\frac{1}{2}-\frac{1-\cos \ell}{\ell^2}\right) \nonumber \\&+&\frac{\Delta}{\mathcal{R}_o \ell^2} \left \langle \delta p_0  [\cos t-\cos(t-\ell)] \right \rangle.
\end{eqnarray}
Therefore, with errors of order $\varepsilon^2$, the peristaltic pumping rate becomes
\begin{multline}
\langle Q \rangle = -\frac{\langle\delta p_0\rangle+\varepsilon \langle\delta p_1\rangle}{\mathcal{R}_o \ell}+ \varepsilon \Delta \left(\frac{1}{2}-\frac{1-\cos \ell}{\ell^2}\right. \\ + \left. \left \langle \frac{\delta p_0}{\mathcal{R}_o\ell^2} [\cos t-\cos(t-\ell)]\right\rangle \right). 
\label{langleQrangle}
\end{multline}
The first contribution corresponds to the familiar Poiseuille-like volumetric rate found by application of a steady pressure gradient. The remaining terms, proportional to $\varepsilon \Delta$, represent the peristaltic effect, which includes a contribution arising from the interaction of the peristaltic wave with the pressure field, which will be seen to be important in the computation of multi-branch networks.

\subsection{The flow in perivascular networks}
\label{sec:theory:network}

A perivascular tree is composed of multiple elements connecting at bifurcating junctions. Consider a general bifurcating tree originating at a single element (or trunk). If the network has $N$ junctions, then the total number of elements is $1+2N$, $1+N$ of which are end elements (or twigs). Application of continuity at the $N$ junctions provides $N$ algebraic equations for the pressure jumps $\delta p_n(t)$ across the $1+2N$ elements. The remaining $1+N$ equations needed to determine $\delta p_n(t)$ are obtained from the known values of the pressure at the $1+N$ downstream ends (measured relative to that at the upstream end of the network). Thus, for each end element, the sum of the pressure jumps $\delta p_n(t)$ along its corresponding trunk-to-twig path must be equal to the pressure at its downstream open end.

In the analysis, the peristaltic wave is assumed to proceed with negligible reflections at the junctions, so that the general expression $r_a=1+\varepsilon \sin (x-t)$ describes the deformation of the artery, with $x$ measuring the distance from the network upstream end (the entrance of the trunk element). For a given element $n$ extending between $x=x_n$ and $x=x_n+\ell_n$ the flow-rate expressions~\eqref{Q_0} and~\eqref{Q_1} take the form
\begin{align} \label{Q_0n}
Q_{0_n}&=-\frac{\delta p_{0_n}}{\mathcal{R}_n \ell_n} +\frac{1}{\ell_n}[\cos(t-x_n)-\cos(t-x_n-\ell_n)] \nonumber \\ &+\sin(t-x_n-\hat{x})
\end{align}
and
\begin{eqnarray} \label{Q_1n}
\langle Q_{1_n} \rangle&=&-\frac{\langle\delta p_1 \rangle_n}{\mathcal{R}_n \ell_n}+\Delta_n \left(\frac{1}{2}-\frac{1-\cos \ell_n}{\ell_n^2}\right) \nonumber \\&+&\frac{\Delta_n}{\ell_n^2} \left \langle \frac{\delta p_{0_n}}{\mathcal{R}_n} [\cos(t-x_n)-\cos(t-x_n-\ell_n)] \right \rangle,
\end{eqnarray}
where $\hat{x}=x-x_n$ in~\eqref{Q_0n}. 

If the time-averaged value of the pressure at the downstream ends differs from the value at the entrance, then $\langle \delta p_{0}\rangle_n \ne 0$, so that the flow rate along each peristaltic element reduces to the trivial Poiseuille solution $\langle Q_{0_n} \rangle_n=-\langle \delta p_{0}\rangle_n/(\mathcal{R}_n \ell_n)$, with the peristaltic contribution being a factor $\varepsilon$ smaller. We focus on configurations with $\langle \delta p_{0} \rangle_n=0$, for which, at leading order in the limit $\varepsilon \ll 1$, the pressure difference across the element and its associated flow rate are harmonic functions that can be expressed in the general form
\begin{equation} 
\label{p0n}
\delta p_{0_n}={\rm Re}\left(P_n  e^{{\rm i}(t-x_n)}\right),
\end{equation}
and
\begin{equation}\label{Q_0n2}
Q_{0_n}={\rm Re}\left[e^{{\rm i}(t-x_n)} \left(-\frac{P_n}{\mathcal{R}_n \ell_n}+\frac{1-e^{-{\rm i}\ell_n}}{\ell_n}- {\rm i}e^{-{\rm i} \hat{x}} \right) \right],
\end{equation}
where $P_n$ are complex numbers, to be determined as part of the solution. On the other hand, using~\eqref{p0n} in~\eqref{Q_1n} gives
\begin{eqnarray} 
\label{Q_1n2}
\langle Q_{1_n} \rangle&=&-\frac{\langle\delta p_1 \rangle_n}{\mathcal{R}_n \ell_n}+\Delta_n \left(\frac{1}{2}-\frac{1-\cos \ell_n}{\ell_n^2}\right) \nonumber \\&+&\frac{\Delta_n}{2 \ell_n^2 \mathcal{R}_n} {\rm Re}\left[P_n (1-e^{{\rm i}\ell_n}) \right],
\end{eqnarray}
which determines the mean flow rate $\langle Q_n \rangle=\varepsilon \langle Q_{1_n} \rangle$. The first term in~\eqref{Q_1n2} is the steady flow rate induced by the steady pressure difference established between the two ends of the element, while the third term is the steady flow rate induced by the interplay of the pressure fluctuations with the peristaltic wave.

The computation of $P_n$ and $\langle\delta p_1 \rangle_n$, whose values are needed in evaluating~\eqref{Q_0n2} and~\eqref{Q_1n2}, requires consideration of mass conservation at the different junctions. Consider the flow at a given junction involving arteries with different radius. Let $n=i$ denote the mother element, whose radius is $r'_{o_i}$, and $n=j$ and $n=k$ denote the daughter elements, with corresponding arterial radii $r'_{o_j}$ and $r'_{o_k}$, respectively. The instantaneous flow rate at the exit of element $i$, given by $Q'_{0_i}=(2\pi \varepsilon \omega/k) r^{\prime 2}_{o_i} Q_{0_i}$ with
\begin{equation}\label{Q_0i}
Q_{0_i}={\rm Re}\left[e^{{\rm i}(t-x_i)} \left(-\frac{P_i}{\mathcal{R}_i \ell_i}+\frac{1-e^{-{\rm i}\ell_i}}{\ell_i}- {\rm i}e^{-{\rm i} \ell_i} \right) \right],
\end{equation}
must be equal to the sum of the flow rates at the entrance of the daughter elements, given by $Q'_{0_j}+Q'_{0_k}=(2\pi \varepsilon \omega/k)(r^{\prime 2}_{o_j}Q_{0_j} + r^{\prime 2}_{o_k}Q_{0_k})$ with
\begin{equation}\label{Q_0j}
Q_{0_j}={\rm Re}\left[e^{{\rm i}(t-x_i-\ell_i)} \left(-\frac{P_j}{\mathcal{R}_j \ell_j}+\frac{1-e^{-{\rm i}\ell_j}}{\ell_j}- {\rm i} \right) \right]
\end{equation}
and
\begin{equation}\label{Q_0k}
Q_{0_k}={\rm Re}\left[e^{{\rm i}(t-x_i-\ell_i)} \left(-\frac{P_k}{\mathcal{R}_k \ell_k}+\frac{1-e^{-{\rm i}\ell_k}}{\ell_k}- {\rm i} \right) \right].
\end{equation}
Substituting~\eqref{Q_0i}--\eqref{Q_0k} into the resulting equation $r^{\prime 2}_{o_i} Q_{0_i}=r^{\prime 2}_{o_j} Q_{0_j}+r^{\prime 2}_{o_k} Q_{0_k}$ leads to
\begin{widetext}
\begin{equation}
\frac{e^{{\rm i}\ell_i} \gamma_i P_i}{\mathcal{R}_i \ell_i}-\frac{\gamma_j P_j}{\mathcal{R}_j \ell_j}-\frac{\gamma_k P_k}{\mathcal{R}_k \ell_k}=\gamma_i \frac{e^{{\rm i}\ell_i}-1}{\ell_i}-\gamma_j\frac{1-e^{-{\rm i}\ell_j}}{\ell_j}-\gamma_k \frac{1-e^{-{\rm i}\ell_k}}{\ell_k}+{\rm i} (\gamma_j+\gamma_k-\gamma_i) \label{PiPjPk}
\end{equation}
\end{widetext}
relating the three pressure constants $P_i$, $P_j$, and $P_k$. In the above equations the arterial radius $r'_{o_n}$ of a given element $n$ has been scaled with the value corresponding to the initial element $n=1$ to give the dimensionless factors $\gamma_n=(r'_{o_n}/r'_{o_1})^2$, which enter as additional parameters in the description. Similarly, using $r^{\prime 2}_{o_i} \langle Q_{1_i} \rangle=r^{\prime 2}_{o_j} \langle Q_{1_j} \rangle+r^{\prime 2}_{o_k} \langle Q_{1_k} \rangle$ for the time-averaged flow rate provides
\begin{widetext}
\begin{align}
\frac{\gamma_i \langle\delta p_1 \rangle_i}{\mathcal{R}_i \ell_i}-\frac{\gamma_j \langle\delta p_1 \rangle_j}{\mathcal{R}_j \ell_j}&-\frac{\gamma_k \langle\delta p_1 \rangle_k}{\mathcal{R}_k \ell_k}=\gamma_i \Delta_i \left(\frac{1}{2}-\frac{1-\cos \ell_i}{\ell_i^2}+ {\rm Re}\left[\frac{P_i (1-e^{{\rm i}\ell_i})}{2 \ell_i^2 \mathcal{R}_i} \right]\right) \nonumber \\
&-\gamma_j \Delta_j \left(\frac{1}{2}-\frac{1-\cos \ell_j}{\ell_j^2}+ {\rm Re}\left[\frac{P_j (1-e^{{\rm i}\ell_j})}{2 \ell_j^2 \mathcal{R}_j} \right]\right)-\gamma_k \Delta_k \left(\frac{1}{2}-\frac{1-\cos \ell_k}{\ell_k^2}+ {\rm Re}\left[\frac{P_k (1-e^{{\rm i}\ell_k})}{2 \ell_k^2 \mathcal{R}_k}  \right]\right) \label{pipjpk}
\end{align}
\end{widetext}
as a relation between $\langle\delta p_1 \rangle_i$, $\langle\delta p_1 \rangle_j$, and $\langle\delta p_1 \rangle_k$.

Equations~\eqref{PiPjPk} and~\eqref{pipjpk}, which apply at the $N$ bifurcations, must be supplemented with the equations stating the known value of the pressure at the downstream ends of the network. Since the pressure jump across the element $\delta p'$ is scaled according to $\delta p=(\delta p')/[2 \pi\mu \varepsilon \omega/(k r'_o)^2]$, the weighting factors $\gamma_n=(r'_{o_n}/r'_{o_1})^2$ enter in the corresponding equations. If, for instance, the pressure at the downstream ends is assumed to be equal to the pressure at the upstream end of the tree, then the equations for each one of the open ends take the form
\begin{equation}
\sum_{j} e^{{\rm i} x_j} P_j/\gamma_j=0 \label{pressure_eqsP}
\end{equation}
and 
\begin{equation}
\sum_{j} \langle \delta p_1 \rangle_j/\gamma_j=0, \label{pressure_eqsp1}
\end{equation}
where the sums include all of the elements contained in the corresponding trunk-to-twig path.

For a given network with $N$ junctions, $1+2N$ elements and $1+N$ downstream ends, the values of the pressure-jump amplitudes $P_n$ ($n=1, \dots, 1+2N$), which are complex numbers, are obtained from a system of linear equations comprising the $N$ continuity-balance equations~\eqref{PiPjPk} along with the $1+N$ pressure equations~\eqref{pressure_eqsP}. Similarly, combining the $N$ equations~\eqref{pipjpk} with the $1+N$ equations~\eqref{pressure_eqsp1} provides a system of $1+2N$ equations for $\langle \delta p_1\rangle_n$. Once the two linear systems are solved, the values of $P_n$ and $\langle \delta p_1\rangle_n$ can be used in~\eqref{Q_1n2} to evaluate the dimensionless flow rate $\langle Q_{1_n} \rangle$ corresponding to each element, which can be expressed in the dimensional form $\langle Q'_{n} \rangle=(2\pi \varepsilon^2 \omega/k) r^{\prime 2}_{o_n} \langle Q_{1_n} \rangle$. A computational algorithm was developed using Python and NumPy~\cite{harris2020array} to solve the problem delineated above; the code is openly available at \href{https://github.com/scientificcomputing/perivascular-peristalsis}{https://github.com/scientificcomputing/perivascular-peristalsis}.

\subsection{Perivascular elements placed in tandem}
\label{sec:theory:tandem}

To illustrate the complications arising in the presence of multiple branches, let us first consider two perivascular elements placed in a tandem arrangement, representing a periarterial space that undergoes a sudden change in cross section. Properties in the two elements will be denoted by the subscripts $a$ and $b$, respectively.

To focus more directly on the peristaltic motion, we shall assume that the pressure takes the same value at the ends $x=0$ and $x=\ell_a+\ell_b$, so that the flow rate can be evaluated in terms of the overpressure at the junction written in the form $p(t)=p_0+\varepsilon p_1+\cdots$. Evaluating~\eqref{Q_0} provides the leading-order expressions
\begin{equation} \nonumber
Q_a=-\frac{p_0}{\mathcal{R}_a \ell_a}+\frac{\cos t-\cos(t-\ell_a)}{\ell_a}+\sin(t-x)
\end{equation}
and
\begin{equation} 
Q_b=\frac{p_0}{\mathcal{R}_b \ell_b}+\frac{\cos(t-\ell_a)-\cos(t-\ell_a-\ell_b)}{\ell_b}+\sin(t-x), \nonumber
\end{equation}
the latter exhibiting a phase lag associated with the propagation of the peristaltic wave along the arterial walls. Equating the two values at the junction ($x=\ell_a$) yields
\begin{align}
p_0&=\left(\frac{1}{\mathcal{R}_a \ell_a}+\frac{1}{\mathcal{R}_b \ell_b}\right)^{-1} \nonumber \\
& \times {\rm Re}\left\{e^{{\rm i}t} \left[\frac{1-e^{-{\rm i}\ell_a}}{\ell_a}-\frac{e^{-{\rm i}\ell_a}(1-e^{-{\rm i}\ell_b})}{\ell_b} \right]\right\},
\end{align}
with ${\rm Re}$ representing the real part of a complex function. Note that, since the overpressure at this order is harmonic, its average value is identically zero, i.e. $\langle p_0\rangle=0$.   

The time-averaged value of the flow rate $\langle Q \rangle$ can be computed from~\eqref{langleQrangle} to give
\begin{widetext}
\begin{align}
\frac{\langle Q \rangle}{\varepsilon} &= -\frac{\langle p_1\rangle}{\mathcal{R}_a \ell_a}+ \Delta_a \left(\frac{1}{2}-\frac{1-\cos \ell_a}{\ell_a^2}+ \frac{1}{\mathcal{R}_a \ell_a^2}  \left \langle p_0 [\cos t-\cos(t-\ell_a)]\right\rangle \right) \nonumber \\ &= \frac{\langle p_1\rangle}{\mathcal{R}_b \ell_b }+ \Delta_b \left(\frac{1}{2}-\frac{1-\cos \ell_b}{\ell_b^2}- \frac{1}{\mathcal{R}_b \ell_b^2}  \left \langle p_0 [\cos(t-\ell_a)-\cos(t-\ell_a-\ell_b)]\right\rangle \right),
\end{align}
where
\begin{equation}
\left \langle p_0 [\cos t-\cos(t-\ell_a)]\right\rangle=\frac{1}{2} \left(\frac{1}{\mathcal{R}_a \ell_a}+\frac{1}{\mathcal{R}_b \ell_b}\right)^{-1} \left[\frac{2(1-\cos \ell_a)}{\ell_a}+\frac{1-\cos \ell_a-\cos \ell_b+\cos(\ell_a+\ell_b)}{\ell_b}\right] 
\end{equation}
and
\begin{equation}
\left \langle p_0 [\cos(t-\ell_a)-\cos(t-\ell_a-\ell_b)]\right\rangle=-\frac{1}{2} \left(\frac{1}{\mathcal{R}_a \ell_a}+\frac{1}{\mathcal{R}_b \ell_b}\right)^{-1} \left[\frac{2(1-\cos \ell_b)}{\ell_b}+\frac{1-\cos \ell_a-\cos \ell_b+\cos(\ell_a+\ell_b)}{\ell_a}\right] 
\end{equation}
Eliminating $\langle p_1\rangle$ and solving for the flow rate finally gives
\begin{align}
\frac{\langle Q \rangle}{\varepsilon} &=\frac{\Delta_a \mathcal{R}_a \ell_a+\Delta_b \mathcal{R}_b \ell_b}{2(\mathcal{R}_a \ell_a+\mathcal{R}_b \ell_b)}-\frac{\Delta_a \mathcal{R}_a^2(1-\cos \ell_a)+\Delta_b \mathcal{R}_b^2(1-\cos \ell_b)}{(\mathcal{R}_a \ell_a+\mathcal{R}_b \ell_b)^2} \nonumber \\ &+\frac{(\Delta_a+\Delta_b)\mathcal{R}_a\mathcal{R}_b}{2(\mathcal{R}_a \ell_a+\mathcal{R}_b \ell_b)^2}[1-\cos \ell_a-\cos \ell_b+\cos(\ell_a+\ell_b)]. \label{eq:tandemflow}
\end{align}
\end{widetext}

\begin{figure}
\begin{center}
\includegraphics[width=0.49\textwidth]{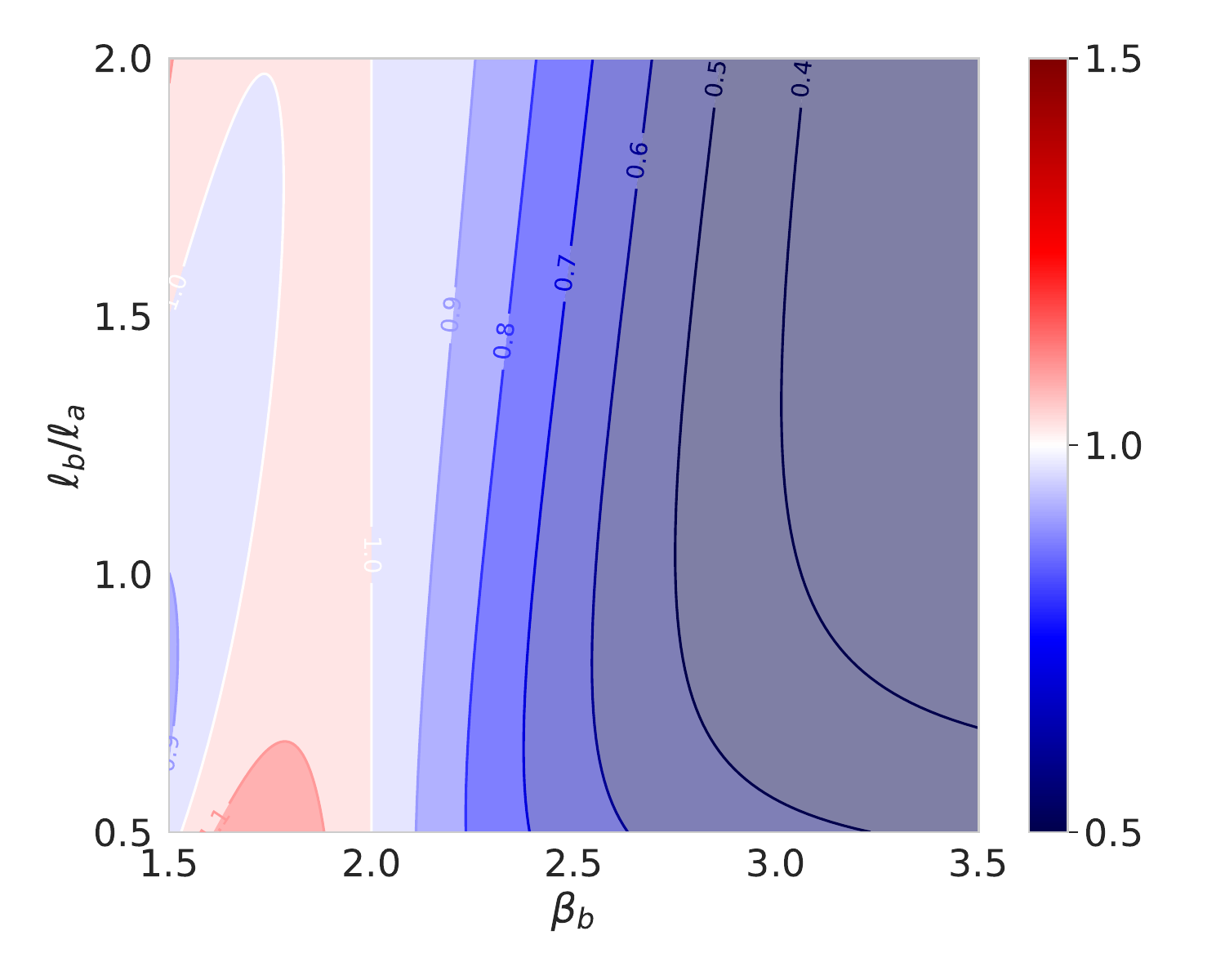}
\includegraphics[width=0.49\textwidth]{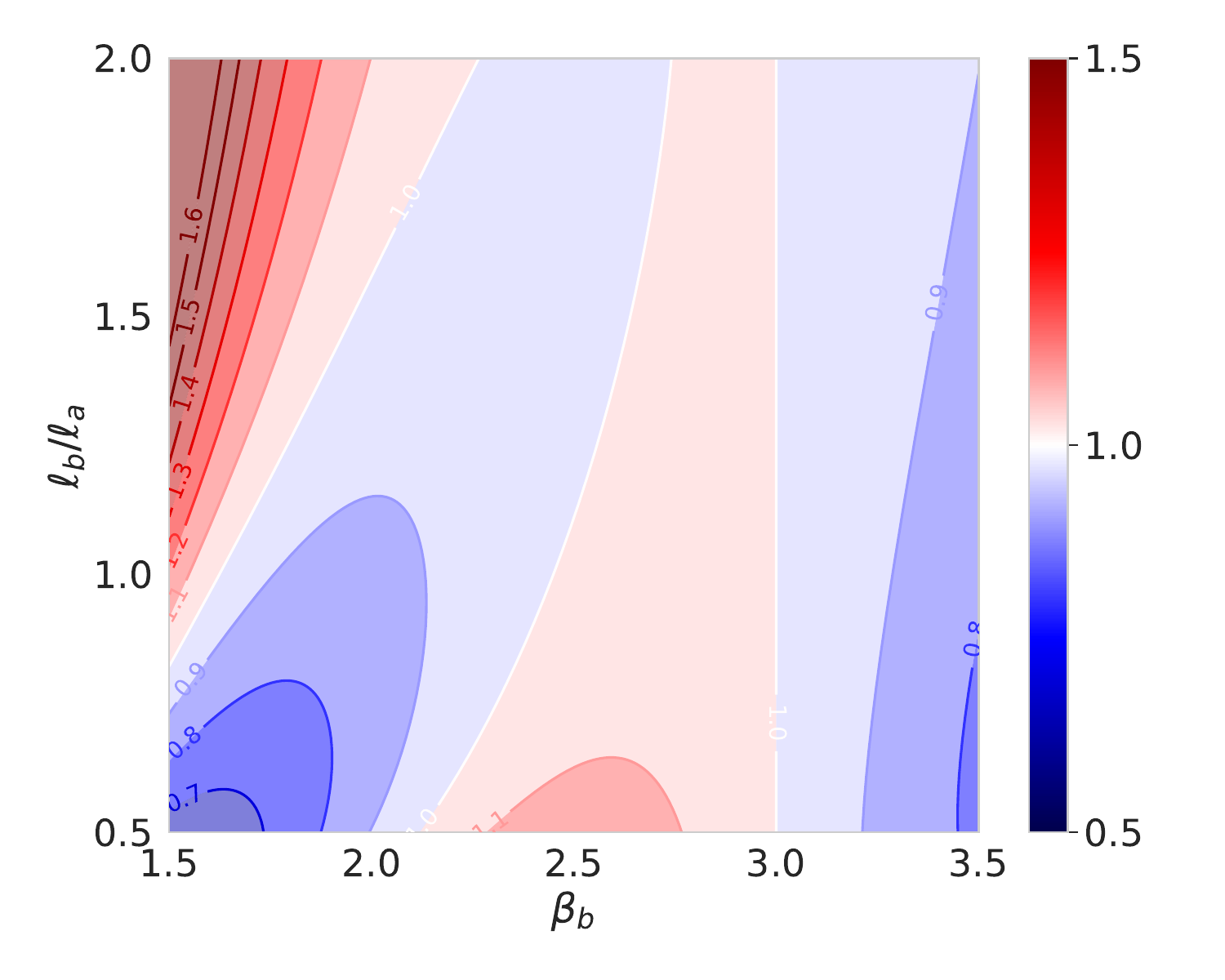}
\end{center}
\caption{The variation with $\ell_b/\ell_a$ and $\beta_b$ of the normalized net flow rate $\langle Q \rangle/\langle Q_{\rm ref} \rangle$ determined from~\eqref{eq:tandemflow} in a tandem configuration with $\ell_a+\ell_b=1$ for $\beta_a = 2.0$ (upper plot) and $\beta_a = 3.0$ (lower plot), with the reference value $\langle Q_{\rm ref} \rangle$ corresponding to the case $\beta_a=\beta_b$.}
\label{fig:tandem-vessel-relflow1}
\end{figure}

As expected, when both elements have the same cross-section, so that $\mathcal{R}_a=\mathcal{R}_b$ and $\Delta_a=\Delta_b=\Delta$, the above expression reduces to
\begin{equation}
\langle Q \rangle=\varepsilon \Delta \left(\frac{1}{2}-\frac{1-\cos(\ell_a+\ell_b)}{(\ell_a+\ell_b)^2}\right),
\label{eq:tandemflow2}
\end{equation}
as corresponds to a peristaltic element of length $\ell_a+\ell_b$ with no imposed pressure difference between its ends; see~\eqref{langleQrangle}. In the general case $\mathcal{R}_a \ne \mathcal{R}_b$ and $\Delta_a \ne \Delta_b$, the element lengths and their hydraulic-resistance parameters enter in a nontrivial way in the computation of the resulting flow rate. Results corresponding to the concentric annular cross section, for which $\mathcal{R}$ and $\Delta$ depend only on the outer-to-inner radius ratio $\beta=r'_e/r'_o$, are shown in Fig.~\ref{fig:tandem-vessel-relflow1}. The color contours represent the relative variation of the flow rate when the radius of the downstream element $\beta_b$ is either smaller or larger than that of the upstream element. In the range of $\beta_b$ represented, the results indicate that when the downstream element has a larger radius (i.e $\beta_b>\beta_a$), corresponding to positive tapering, the flow rate decreases. The results for $\beta_b<\beta_a$ (negative tapering) reveal a more complicated non-monotonic behavior, with the variation of $\langle Q \rangle$ with decreasing $\beta_b$ for a given ratio $\ell_b/\ell_a$ exhibiting a local maximum followed by a local minimum. 

\begin{figure}
\begin{center}
\includegraphics[width=0.48\textwidth]{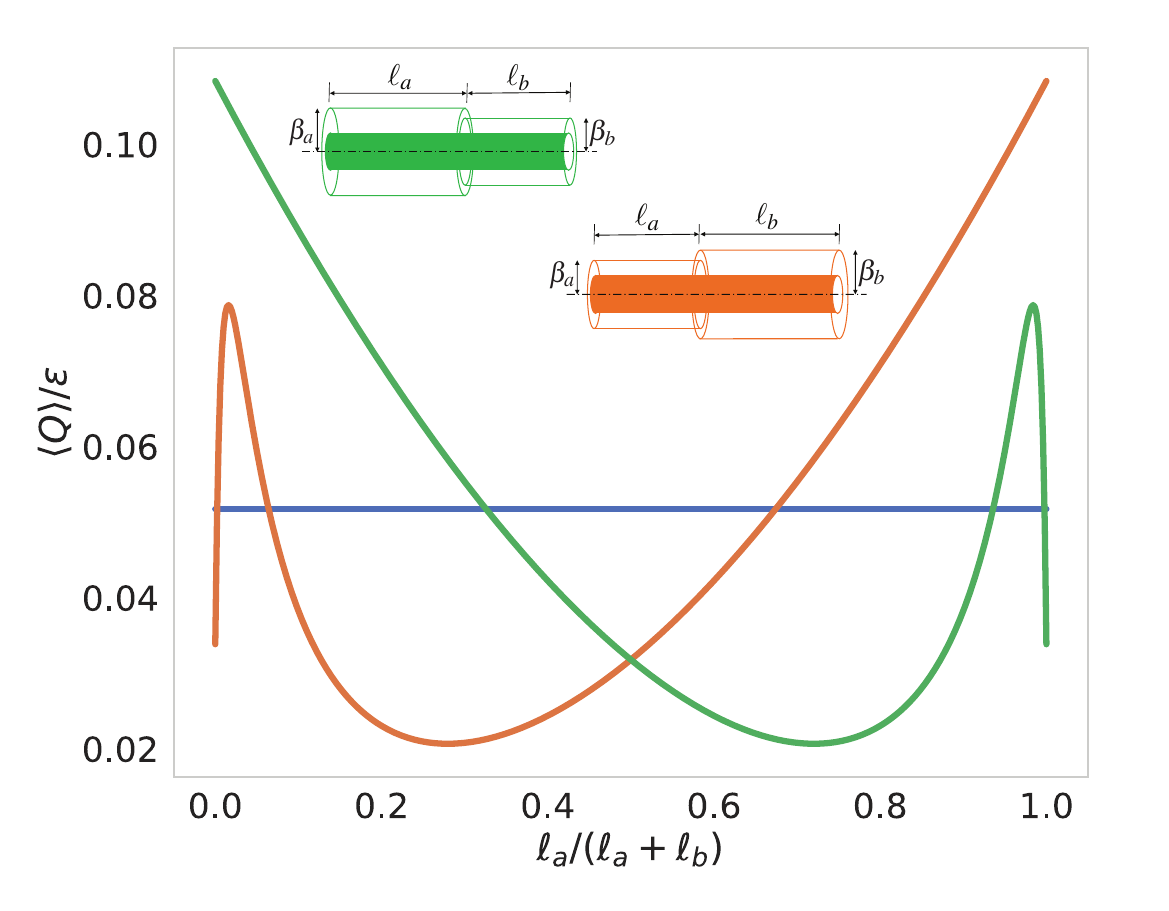}
\end{center}
\caption{The variation with $\ell_a/(\ell_a+\ell_b)$ of the reduced flow rate $\langle Q \rangle/\varepsilon$ as obtained from~\eqref{eq:tandemflow} for a tandem configuration with $\ell_a+\ell_b=1$ and $\beta_a = \beta_b = 3.0$ (blue), $\beta_a = 2.0, \beta_b = 4.0$ (orange), and $\beta_a = 4.0, \beta_b = 2.0$ (green).}
\label{fig:tandem-vessel-relflow2}
\end{figure}

The non-monotonicity noticed above has interesting implications that are further investigated in Fig.~\ref{fig:tandem-vessel-relflow2} using two elements in tandem with $\ell_a+\ell_b=1$. The blue line corresponds to two identical elements with $\beta_b=\beta_a=3$, for which $\langle Q \rangle/\varepsilon=0.0517$, as follows from~\eqref{eq:tandemflow2},  while the orange and green curves correspond to a two-element system with positive ($\beta_a=2<\beta_b=4$) and negative ($\beta_a=4>\beta_b=2$) tapering, respectively. As can be seen, depending on the relative length of the elements, both positive and negative tapering can result in flow rates that are either larger or smaller than that of the constant-cross-section configuration. The orange and green curves are symmetric about $\ell_a/(\ell_a+\ell_b)=0.5$, reflecting the symmetry present in~\eqref{eq:tandemflow}. This implies that, for two different elements, the resulting flow rate is independent of the order in which they are arranged. The curves show a strong non-monotonicity, consistent with the results shown in Fig.~\ref{fig:tandem-vessel-relflow1}. Focusing on the orange curve, one can see that as $\ell_a$ increases from $\ell_a=0$ ($\ell_b=1$) to $\ell_a=1$ ($\ell_b=0$) the flow rate transitions from the value $\langle Q \rangle/\varepsilon=0.0338$ corresponding to a uniform element with $\beta=\beta_b=4$ to the much larger value  $\langle Q \rangle/\varepsilon=0.1084$ corresponding to a uniform element with $\beta=\beta_a=2$. Interestingly, there is a range of values of $\ell_a$ for which the resulting flow rate is smaller than both limiting values, indicating that tapering may, in some cases, have a detrimental effect on the induced flow rate. The fact that this unexpected behavior arises in this simple network configuration underscores the level of complexity to be expected from the nonlinear interactions arising in multi-branch peristaltic systems. While the tandem configuration affords a simple explicit closed-form expression for the flow rate, in the general case the computation involves the solution to a system of linear equations, delineated above in Section~\ref{sec:theory:network}.

\subsection{Analysis of a single-bifurcation network}

It is instructive to consider a peristaltic network with a single bifurcation connecting a mother element $a$ with two daughter elements $b$ and $c$, all arteries having the same radius $r'_o$, so that $\gamma_a=\gamma_b=\gamma_c$. The upstream and downstream ends are assumed to be at the same pressure, so that $\delta p_a=-\delta p_b=-\delta p_c$, where the pressure jump across the mother element $\delta p_a$ is also the instantaneous value of the overpressure at the junction. From~\eqref{p0n} it then follows that $P_a=-P_b e^{-\ell_a {\rm i}}=-P_c e^{-\ell_a {\rm i}}$, which can be substituted into~\eqref{PiPjPk} to give
\begin{widetext}
\begin{equation}
e^{\ell_a {\rm i}} P_a=-P_b=-P_c=\frac{(e^{{\rm i}\ell_a}-1)/\ell_a + {\rm i} -(1-e^{-{\rm i}\ell_b})/\ell_b-(1-e^{-{\rm i}\ell_c})/\ell_c}{1/(\mathcal{R}_a \ell_a)+1/(\mathcal{R}_b \ell_b)+1/(\mathcal{R}_c \ell_c)},\label{PaPbPc}
\end{equation}
with the associated value of the pressure at the bifurcation $p_0=\delta p_{0_a}=-\delta p_{0_b}=-\delta p_{0_c}={\rm Re}(P_a e^{{\rm i}t})$ reducing to 
\begin{equation}
p_{0}=\left(\frac{1}{\mathcal{R}_a \ell_a}+\frac{1}{\mathcal{R}_b \ell_b}+\frac{1}{\mathcal{R}_c \ell_c}\right)^{-1} \times \left[\frac{\cos t}{\ell_a}-\left(\frac{1}{\ell_a}+\frac{1}{\ell_b}+\frac{1}{\ell_c}\right) \cos(t-\ell_a)+\frac{\cos(t-\ell_a-\ell_b)}{\ell_b}+\frac{\cos(t-\ell_a-\ell_c}{\ell_c}-\sin(t-\ell_a) \right].
\end{equation}
Similarly, the steady pressure differences across the different elements can be evaluated from~\eqref{pipjpk} to give
\begin{align}
\langle\delta p_1 \rangle_a&=-\langle\delta p_1 \rangle_b=-\langle\delta p_1 \rangle_c=\left(\frac{1}{\mathcal{R}_a \ell_a}+\frac{1}{\mathcal{R}_b \ell_b}+\frac{1}{\mathcal{R}_c \ell_c}\right)^{-1} \times \left\{\Delta_a \left(\frac{1}{2}-\frac{1-\cos \ell_a}{\ell_a^2}+ {\rm Re}\left[\frac{P_a (1-e^{{\rm i}\ell_a})}{2 \ell_a^2 \mathcal{R}_a} \right]\right) \right.\nonumber \\
&\left.-\Delta_b \left(\frac{1}{2}-\frac{1-\cos \ell_b}{\ell_b^2}+ {\rm Re}\left[\frac{P_b (1-e^{{\rm i}\ell_b})}{2 \ell_b^2 \mathcal{R}_b} \right]\right)-\Delta_c \left(\frac{1}{2}-\frac{1-\cos \ell_c}{\ell_c^2}+ {\rm Re}\left[\frac{P_c (1-e^{{\rm i}\ell_c})}{2 \ell_c^2 \mathcal{R}_c} \right]\right)\right\}. \label{papbpc}
\end{align}
\end{widetext}
The expressions given in~\eqref{PaPbPc} and~\eqref{papbpc} can be used in~\eqref{Q_1n2} to determine the steady flow rates $\langle Q\rangle=\varepsilon \langle Q_1\rangle$ circulating along each element, with $\langle Q_a\rangle=\langle Q_b\rangle+\langle Q_c\rangle$, as follows from~\eqref{papbpc}. The predicted values are to be compared with numerical results in the next section.

\section{Numerical simulations of peristaltic pumping}
\label{sec:numerics}

As summarized in Section \ref{sec:problem}, fluid flow due to peristaltic pumping can be modelled as Stokes flow through an annular domain, where the inner boundary deforms in time. Thus far, we have used lubrication theory to dissect the governing forces for this flow. In this section, we show how the original model can be discretized using a reduced-order finite element method, suitable for direct numerical simulations of cross-section average pressure and flux fields. We start by giving the network model and its finite element discretization (Section~\ref{sec:numerics:fem}), before evaluating the accuracy of the analytically derived flow rate estimates by comparison with the numerical counterparts for three different configurations: the single vessel, the tandem vessel and a simple bifurcation (Section~\ref{sec:numerics:validation}).

\subsection{Network model and discretization}
\label{sec:numerics:fem}

The numerical simulations are performed on a network model describing peristaltic pumping as previously developed~\cite{gjerde2023graphnics}. Let $G=(V,E)$ denote a directed graph representing the perivascular network, with edges $E$ and nodes $V$. Each edge $e_i \in E$ is associated with a centerline $\Lambda_i'$ for the vessel, and each node $v_j \in V$ is associated with a spatial point $x_j'$. 

We model fluid flow through the perivascular network via the three-dimensional Stokes flow equations
\begin{gather}
\begin{aligned}
\frac{\partial \mathrm{v}'}{\partial t'} + \nu \nabla^2 \mathrm{v}' + \nabla p' &= 0 && \text{  in  }  \Omega_i' \times (0,T], \\
\nabla \cdot \mathrm{v}' &= 0 && \text{  in  }  \Omega_i' \times (0,T],  \\
\mathrm{v}' &= \mathrm{v}'_0 &&\text{ at }  \Omega_i' \times \{ t'=0 \}, \label{eq:stokes}
\end{aligned}
\end{gather}
where $\Omega_i'$ denotes the perivascular domain for branch $i$ in the perivascular network, $\nu$ denotes the kinematic viscosity, $\mathrm{v}'=(u',v',w')$ is the velocity, whose initial value is $\mathrm{v}_0'$, and $T$ denotes the simulation end time. We take $T=T_{\text{cyc}} \cdot n_{\text{cyc}}$, for an integer number of cycles $n_{\text{cyc}}$ and where $T_{\text{cyc}}=2\pi/\omega$ denotes the period of the travelling wave. 

The full model \eqref{eq:stokes} can be reduced topologically to one dimension by integrating over the two-dimensional cross-section of the domain. Doing this for the conservation of mass equation yields its network equivalent in terms of the cross-section flux $Q'$:
\begin{align}
    \frac{\partial Q'}{\partial x'} = \frac{\partial A'}{\partial t'} \text{ on } \Lambda_i' \label{eq:cons-mass}
\end{align}
where $A^{'}$ denotes the cross-section area. This expression is equivalent to the one given in \eqref{cont_int}. Next, one can integrate the momentum equation in \eqref{eq:stokes}, which yields its network equivalent
\begin{align}
  \frac{\partial Q'}{\partial t'} + \frac{\nu}{A'}  \frac{\partial^2 Q'}{\partial x'^2} +	\mathcal{R}' Q' + \frac{\partial p'}{\partial x'} = 0 \text{ on } \Lambda_i', \label{eq:cons-mom}
\end{align}
where $p'$ denotes the average pressure at each cross-section and $\mathcal{R}'$ is the dimensional resistance $\mathcal{R}'=\mathcal{R}/r_0'^4$, as derived by Tithof et al~\cite{tithof2019hydraulic}. As noted above, $ \partial^2 Q'/\partial x'^2$, is typically vanishingly small for perivascular flows; we therefore neglect it.

Finally, we join together these equations at each bifurcation, by asking for pressure continuity and conservation of mass at the bifurcation. The latter reads:
\begin{align}
    \sum_{e_i \in E_{\text{in}}(v_j)} Q_i' (x_j')= \sum_{e_i \in E_{\text{out}}(v_j)} Q_i' (x_j') \label{eq:bif}
\end{align}
where we use $E_{\text{in}}(v_j)$ and $E_{\text{out}}(v_j)$ to denote vessels going into and out node $v_j$ with coordinate $x_j'$, respectively. Equations \eqref{eq:cons-mass}-\eqref{eq:bif} comprise the reduced model; it has previously been validated against the full model by Daversin--Catty et al~\cite{daversin2022geometrically}.

Let $\Lambda' = \cup_{i=1}^n \Lambda_i'$ denote the network domain and consider a discrete mesh $\Lambda_h$ of $\Lambda'$ with characteristic mesh cell size $h$. We define a set of discrete finite element spaces $V_h$ and $M_h$ relative to $\Lambda_h$. Specifically, we set
\begin{itemize}
  \item $V_h$ to be the space of (discontinuous) piecewise constants;
  \item $M_h$ to be the space of continuous piecewise linears.
\end{itemize}
The reduced equations~\eqref{eq:cons-mom}-\eqref{eq:bif} can then be expressed in variational form as: find $Q'_{h} \in V_h$, $p'_{h} \in M_h$ such that
\begin{gather}
\begin{aligned}
    \left( \partial_{t'} Q'_{h}+ \mathcal{R}' Q'_{h}, \psi  \right)_{\Lambda'} +  \left( \partial_{x'} p'_{h}, \psi  \right)_{\Lambda'} &=  0 \\
    \left( \partial_{x'} Q'_{h}, \phi  \right)_{\Lambda'} &= \left( \partial_{t'} A, \phi  \right)_{\Lambda'} \label{eq:disc-eq-space}
\end{aligned}
\end{gather}
for all $\psi \in V_h$, $\phi \in M_h$. This discrete formulation is well-posed, numerically stable and provides mass conservation at bifurcation points.

\begin{widetext}
Finally, we discretize in time using a (first-order) implicit Euler scheme. Let K denote the number of time steps. Let $Q'_{h,0}$ denote a (given) initial flow at $t'=0$. For each time step $k \in [1,.., K]$, we solve the following system: Find $Q'_{h,k+1} \in V_h$, $p'_{h,k+1} \in M_h$ such that
\begin{gather}
\begin{aligned}
    \left(Q'_{h,k+1}+ \Delta t \mathcal{R}' Q'_{h,k+1}, \psi  \right)_{\Lambda'} +   \Delta t \left( \partial_{x'} p'_{h,k+1}, \psi  \right)_{\Lambda'} &=  \left(Q'_{h,k}, \psi  \right)_{\Lambda'} \\
    \left( \partial_{x'} Q'_{h,k+1}, \phi  \right)_{\Lambda'} &= \left( \partial_{t'} A^{'}_{k+1}, \phi  \right)_{\Lambda'} \label{eq:disc-eq}
\end{aligned}
\end{gather}
for all $\psi \in V_h$, $\phi \in M_h$, where $\Delta t=T/K$ denotes the time step size. 
\end{widetext}
Finally, given a flux $Q'$ solving \eqref{eq:disc-eq}, we compute the net flow for the last cycle as
\begin{align*}
\langle Q_h' \rangle = \frac{1}{T_\text{cyc}} \sum_{K-K_{\text{cyc}}}^K \Delta t \, Q_h'(x',t'_k),
\end{align*}
where $K_{\text{cyc}}=K/n_{\text{cyc}}$ denotes the number of time steps per cycle.

\subsection{Numerical validation}
\label{sec:numerics:validation}

We now focus on comparing the numerical $\langle Q_h' \rangle$ and analytical predictions $\langle Q' \rangle$ for the dimensional net flow rates induced by peristaltic pumping. The numerical simulations were performed by solving \eqref{eq:disc-eq} with zero initial cross-section flux ($Q_{h,0}'=0$) with different time steps $\Delta t$. For each simulation, the number of cycles $n_{\text{cyc}}$ was chosen large enough for the system to settle in its periodic state. A decreasing difference (between the numerical and analytical flow rates) with decreasing time step indicates that the difference primarily stems from the numerical approximation. Conversely, a stable difference with respect to the time step suggests that the discrepancy is attributable to the assumptions made when deriving the analytical solution.

Table \ref{tab:validation} provides a comparison between the numerically computed and analytically determined net flow rate values. The table includes three cases: a single vessel with a length of $L=1$mm, a tandem vessel composed of two segments with lengths $L_a=L_b=0.5$mm, and a simple bifurcation consisting of three segments with equal lengths $L_a=L_b=L_c=1$mm. All vessels are assigned radius $r_o'=0.1$mm. Unless otherwise stated, we use $\omega=2\pi$ and $\beta=2$. The differences between the numerical and analytical values were all less than 1.5\% or lower for the smallest time step. Furthermore, it was observed that the discrepancy decreases as the time refinement is increased, indicating improved accuracy and convergence of the numerical solution.

\begin{widetext}

\begin{table}
\caption{Comparison of analytically and numerically computed net flows, i.e. $\langle Q' \rangle$ and $ \langle Q_h' \rangle$, for different PVS configurations.}
\begin{subtable}{0.99\textwidth}

\caption{Single perivascular element, $\epsilon=0.1$}

{$\lambda=1$} \hspace{20em} {$\omega=2\pi$} \vspace{1em}

\begin{tabular}{ l | l l | l } \hline \toprule
   &  $\langle Q_h' \rangle$  &  & {$\langle Q' \rangle$}     \\ 
$\omega/2\pi$    & ${\Delta t=T/25}$ & ${\Delta t=T/50}$ &  \\ \midrule
0.1     &  \footnotesize 8.457e-05 (0.12\%) & \footnotesize 8.447e-05 (0.00\%) &  \footnotesize 8.447e-05\\ 
1       &  \footnotesize 8.457e-04 (0.12\%) & \footnotesize 8.447e-04 (0.00\%) & \footnotesize  8.447e-04 \\ 
10      &  \footnotesize 8.457e-03 (0.12\%) & \footnotesize 8.447e-03 (0.00\%) & \footnotesize  8.447e-03\\ 
100     &  \footnotesize 8.457e-02 (0.12\%) & \footnotesize 8.447e-02 (0.00\%) & \footnotesize 8.447e-02\\ \bottomrule
\end{tabular} \hspace{2em}
\begin{tabular}{ l | l l | l } \hline \toprule
   &  $\langle Q_h' \rangle$  &  & {$\langle Q' \rangle$}     \\ 
{$\lambda$}    & ${\Delta t=T/50}$ & ${\Delta t=T/100}$ &  \\ \midrule
0.1 &  \footnotesize 8.446e-05 (0.02\%) & \footnotesize 8.445e-05 (0.03\%)  & \footnotesize  8.447e-05\\ 
1   &  \footnotesize 8.447e-04 (0.00\%) & \footnotesize 8.446e-04 (0.02\%) & \footnotesize 8.447e-04\\ 
2   & \footnotesize  1.008e-03 (0.31\%) &  \footnotesize 1.005e-03 (0.07\%) & \footnotesize 1.005e-03   \\
10  &  \footnotesize 2.821e-04 (2.84\%) & \footnotesize  2.765e-04 (0.80\%) & \footnotesize 2.743e-04\\  \bottomrule\end{tabular}
\label{tab:validation_single}
\end{subtable}

\vspace{2em}
\begin{subtable}{0.99\textwidth}
\caption{Tandem perivascular elements, $\epsilon=0.1$}

$\beta_a = 2$, $\beta_b = 3$ \hspace{15em} $\beta_a = 4$, $\beta_b = 2$, $L_a=1$, $L_b=2$ \vspace{2em}

\begin{tabular}{ l | l l | l } \hline \toprule
   &  $\langle Q_h' \rangle$  &  & {$\langle Q' \rangle$}     \\ 
{$\lambda$}    & ${\Delta t=T/25}$ & ${\Delta t=T/50}$ &  \\ \midrule
0.1  & \footnotesize 8.077e-05 (0.06\%)  & \footnotesize 8.070e-05 (0.04\%)   & \footnotesize 8.073e-05 \\ 
1    & \footnotesize 8.080e-04 (0.10\%)  & \footnotesize 8.072e-04 (0.01\%)   & \footnotesize 8.073e-04 \\ 
2    & \footnotesize 1.135e-03 (1.59\%)  & \footnotesize 1.126e-03 (0.76\%)   & \footnotesize 1.117e-03 \\ 
10   & \footnotesize 4.707e-04 (6.70\%)  & \footnotesize 4.452e-04 (0.93\%)   & \footnotesize 4.411e-04 \\ 
\end{tabular} \hspace{2em}
\begin{tabular}{ l | l l | l } \hline \toprule
   &  $\langle Q_h' \rangle$  &  & {$\langle Q' \rangle$}     \\ 
{$\lambda$}    & ${\Delta t=T/25}$ & ${\Delta t=T/50}$ &  \\ \midrule
0.1  & \footnotesize 3.895e-05 (0.27\%)   & \footnotesize 3.885e-05 (0.01\%)  & \footnotesize 3.884e-05 \\ 
1    & \footnotesize 3.895e-04 (0.26\%)   &\footnotesize 3.884e-04 (0.01\%)  & \footnotesize 3.884e-04 \\ 
2    & \footnotesize 7.772e-04 (0.35\%)   & \footnotesize 7.741e-04 (0.05\%)  & \footnotesize 7.745e-04 \\ 
10   & \footnotesize 7.824e-04 (13.10\%)  & \footnotesize 7.119e-04 (2.91\%)  & \footnotesize 6.917e-04 \\ 
\end{tabular}
\label{tab:validation_tandem}
\end{subtable}

\vspace{2em}
\begin{subtable}{0.99\textwidth}
\caption{Perivascular bifurcation}
\begin{center}

$\beta_a =3,\beta_b=\beta_c=2$, $L_a=2, L_b=L_c=1$, \hspace{3em}  $\beta_a =2, \beta_b=3, \beta_c=4$, $L_a=1, L_b=2, L_c=3$

\begin{tabular}{ l | l l | l | l } \hline \toprule
   &  $\langle Q_h' \rangle$  &  & {$\langle Q' \rangle$}     \\ 
{$\lambda$}    & ${\Delta t=T/25}$ & ${\Delta t=T/50}$ &  \\ \midrule
0.1  & \footnotesize 1.348e-04 (0.47\%)   & \footnotesize 1.346e-04 (0.35\%) & \footnotesize 1.341e-04 \\ 
1    & \footnotesize 1.347e-03 (0.42\%)   & \footnotesize 1.345e-03 (0.29\%) & \footnotesize 1.341e-03 \\ 
2    & \footnotesize 1.969e-03 (0.75\%)   & \footnotesize 1.952e-03 (0.13\%) & \footnotesize 1.954e-03 \\ 
10   & \footnotesize 1.905e-03 (9.59\%)   & \footnotesize 1.827e-03 (5.10\%) & \footnotesize 1.739e-03 \\ 
\end{tabular}  \hspace{2em}
\begin{tabular}{ l | l l | l | l } \hline \toprule
   &  $\langle Q_h' \rangle$  &  & {$\langle Q' \rangle$}     \\ 
{$\lambda$}    & ${\Delta t=T/25}$ & ${\Delta t=T/50}$ &  \\ \midrule
0.1  & \footnotesize 8.380e-05 (0.17\%) & \footnotesize 8.368e-05 (0.03\%)  & \footnotesize 8.365e-05 \\ 
1  & \footnotesize 8.377e-04 (0.14\%)   & \footnotesize 8.364e-04 (0.01\%)  & \footnotesize 8.365e-04 \\ 
2  & \footnotesize 1.071e-03 (0.87\%)   & \footnotesize 1.061e-03 (0.13\%)  & \footnotesize 1.062e-03 \\ 
10  & \footnotesize 8.263e-04 (2.78\%)  & \footnotesize 7.993e-04 (0.58\%)  & \footnotesize 8.040e-04 \\ 
\end{tabular}  

\end{center}
\label{tab:validation_bif}
\end{subtable}
\label{tab:validation}
\end{table}
\end{widetext}

\section{Net perivascular flow in physiological regimes}

\label{sec:physiological}

Attention is directed now to the question of whether peristaltic pulsations at physiologically relevant scales in time and space can drive net directional flow in perivascular networks, with particular emphasis on cardiac pulse waves, spontaneous vasomotion, and the modulatory effect of sleep.

\subsection{Optimal wave lengths and peak net flow}

To begin to investigate the dependencies of the perivascular net flow on the different physiological parameters, it is illustrative to consider the flow in a single perivascular element of length $L$ in which the pressure takes identical values at both ends (i.e. $p'_0=p'_L$). Then, the net flow rate $\langle Q \rangle$ given by~\eqref{langleQrangle}, written in dimensional form via \eqref{Q_eq}, reduces to
\begin{equation} 
\label{Q'exp}
\langle Q' \rangle = 2\pi \varepsilon^2 \omega r_o'^2 L \Delta \, F(\ell)
\end{equation}
where 
\begin{equation} \label{f_function}
F(\ell)=\frac{1}{\ell}\left(\frac{1}{2}-\frac{1-\cos \ell}{\ell^2}\right)
\end{equation}
with $\ell=2\pi L/\lambda$. Equation~\eqref{Q'exp} reveals that the net flow rate increases linearly with the angular frequency of the travelling wave $\omega$ and quadratically with its amplitude $\varepsilon$ and with the arterial radius $r_o'$. The parameter $\Delta$ accounts for the effect of the cross-sectional perivascular shape, with~\eqref{Delta_eq2} describing a concentric annulus of outer-to-inner radius ratio $\beta = r'_e/r'_o$. As is clear from Fig.~\ref{fig:RDelta}, the flow rate is strongly dependent on this parameter, indicating that an accurate description of the anatomical shape of the perivascular space is fundamental for enabling a precise evaluation of the flow rate.

\begin{figure}
  \begin{center}
    \includegraphics[width=0.45\textwidth]{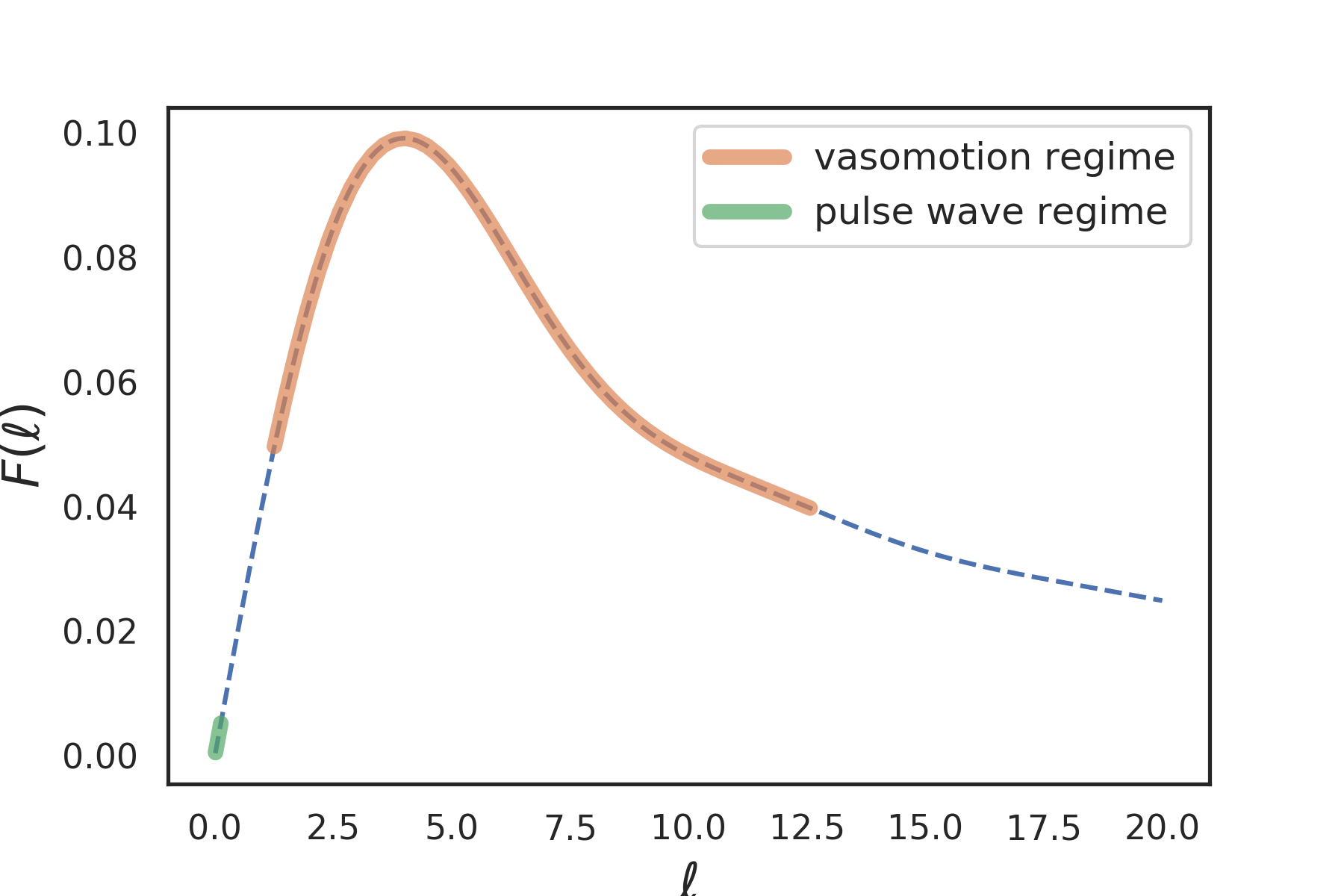}
  \end{center}
  \caption{The net flow in a perivascular segment depends on the ratio of characteristic length $L$ to peristaltic wavelength $\lambda$. The plot shows the scaling function $F$ of the net flow rate \eqref{f_function}. Shaded regions (orange and green) represent estimated regimes for vasomotion and pulse waves (see.~Table~\ref{tab:physiological}). Fractional lengths $\ell=2 \pi L/\lambda$ were computed for vessel lengths $L$ between 1\% and 10\% of the rat brain length $L_{\text{brain}}=20$mm (i.e. $0.01 \, L_{\text{brain}} \le L \le 0.1 \, L_{\text{brain}}$).
  }
  \label{fig:ffunction}
\end{figure}

The function $F(\ell)$, carrying the dependence on the wavelength, is represented in Fig.~\ref{fig:ffunction}. As can be seen, $F$ is non-monotonic and reaches a maximum value $F_{max} \simeq 0.09916$ at $\ell=2\pi L/\lambda \simeq 3.9959$. Thus, for a perivascular element of given length $L$, the maximum net flow rate induceable by the peristaltic wave, achieved when the peristaltic wave has a wavelength $\lambda=1.5724 L$, is
\begin{equation*}
  \langle Q' \rangle_{max} \simeq 0.623 \varepsilon^{2} \omega r_o'^2 L \Delta.
\end{equation*} 
On the other hand, the function $F$ takes simple limiting forms for extreme values of $\ell =2\pi L/\lambda$. For small wavelengths (i.e. large values of $\ell$), the expression~\eqref{f_function} reduces to $F \simeq 1/(2\ell)$, so that 
\begin{equation} \label{Q'exp2}
\langle Q' \rangle = \varepsilon^2 \omega r_o'^2 \Delta \lambda/2 \quad {\rm for} \quad \lambda \ll L,
\end{equation}
independent of $L$, while in the opposite limit $\ell=2\pi L/\lambda \ll 1$ of large wavelengths one finds $F \simeq \ell/24$, yielding
\begin{equation} \label{Q'exp3}
\langle Q' \rangle = \varepsilon^2 \omega r_o'^2 \Delta \frac{(\pi L)^2}{6 \lambda} \quad {\rm for} \quad \lambda \gg L.
\end{equation}

The net flow rate induced by peristaltic pumping thus depends on the ratio between the arterial length $L$ and the peristaltic wave length $\lambda$ through the parameter $\ell =2\pi L/\lambda$. This ratio may differ between physiological regimes, systems, and species. Characteristic values of wave speeds $c$, wavelengths $\lambda$ and frequencies $f=\omega /(2 \pi)$ corresponding to different relevant peristaltic waves are listed in Table~\ref{tab:physiological} along with the optimal arterial length $L_{\rm opt}=3.9959 \lambda/(2\pi)$ for which the corresponding induced flow rate would be maximum. 
With a wavelength of $\lambda \simeq$ 100mm (in mice) and $\lambda \simeq$ 2000 mm (in humans), the cardiac pulse wave would yield near peak net flow rates at scales that are on the order of the respective body scales (Table~\ref{tab:physiological}), but negligibly small net flow at the length scales found in the cerebral periarterial system (Fig.~\ref{fig:ffunction}). In contrast, the peristaltic pumping induced by vasomotion or similar low-frequency waves at wave speeds of 0.1--0.4 mm/s is optimal at millimetric scales. These estimates thus lend support to the hypothesis that vasomotion or similar low frequency/low wave speed wall motions may drive more substantial directional perivascular flow at physiologically relevant scales.
\begin{table}
\vspace{1em}
  \begin{tabular}{ l | r r r r | r}
  \toprule
  Wave type & $c$ (mm/s) & $\lambda$ (mm) &  f (Hz) & L$_{opt}$ (mm) & Ref. \\ 
  \midrule
 {\scriptsize pulse wave (mice)} & 1000 & 100 & 10 & 63.7 & \cite{daversin2020mechanisms} \\
 {\scriptsize pulse wave (human, brain)} & 2380 & 2204 & 1.08 & 1403 & \cite{jung2021novel} \\
 {\scriptsize vasomotion (rat, mesentry)} & 0.1 & 1 & 0.1 & 0.637 &\cite{seppey2010intercellular} \\
 {\scriptsize vasomotion (mice, pial)} & 0.4 & 4 & 0.1 & 2.55 & \cite{munting2023spontaneous} \\
 
 \bottomrule
\end{tabular}
\caption{Physiological travelling wave characteristics -- c: wave speed, $\lambda$: wavelength, $f=\omega/(2\pi)$: frequency ($c = \lambda f$) and $L$ characteristic length. The optimal characteristic length $L_{\rm opt}$ is estimated in order for peristalsis to be at its most effective based on measured wave speeds and frequencies: $L_{\rm opt}=3.9959 \lambda/(2\pi)$.} 
\label{tab:physiological}
\end{table}

\subsection{Vasomotion peristaltic pumping}

To further assess the relevance of vasomotion peristaltic pumping in connection with perivascular flow one may use~\eqref{Q'exp} to estimate the magnitude of the associated flow velocity. Consider as an illustrative example perivascular pumping in the mouse brain for an estimated frequency $f=\omega/(2\pi)=0.1$ Hz assuming for simplicity that the cross section has a concentric annular shape and that the peristaltic wave has a dimensionless amplitude $\varepsilon=0.1$, the latter estimate motivated by experimental reports of spontaneous or stimulus-evoked vasomotion in awake mice~\cite{van2020vasomotion, munting2023spontaneous}. The maximum flow rate $\langle Q' \rangle_{\max} = 4 \pi^2 \Delta \, F_{max} \varepsilon^2 f \, r_o'^2 L_{\rm opt}$ is achieved for a peristaltic element of length $L=L_{\rm opt} \simeq 2550 \mu$m and $F=F_{max} \simeq 0.09916$, thereby yielding
\begin{equation}  \label{umeaneq}
u'_{\rm mean} = \frac{4 \pi \Delta}{\beta^2-1} F_{max} \varepsilon^2 f L_{\rm opt}
\end{equation}
for the corresponding PVS cross-section-averaged velocity $u'_{\rm mean}=\langle Q' \rangle_{\max}/[\pi (r_e^{\prime^2}-r_o^{\prime^2})]$. The resulting value depends strongly on the shape of the cross section, entering through the parameter $\beta$. For instance, if one uses $\beta=3$ in the evaluation of $\Delta=1.283$, a selection that is consistent with the values $r_o' = 20$ $\mu$m and $r_e' = 60$ $\mu$m reported by Mestre et al.\cite{mestre2018flow}, then it follows from~\eqref{umeaneq} that $u'_{\rm mean}=0.51$ $\mu$m/s. Moreover, using values from the recent work of Bojarskaite~\cite{bojarskaite2023sleep}, considering mouse penetrating arterioles with a medium arteriole diameter at baseline of $\approx 12 \mu$m ($r'_o = 6 \mu$m) and PVS width of $\approx 7 \mu$m ($r'_e = 13\mu$m) yields $\beta = 2.16$, and $u'_{\rm mean} = 1.96 \mu$m. In contrast, when the computation considers $\beta=1.5$ ($\Delta=5.613$), corresponding to the values $r_o' = 50$ $\mu$m and $r_e' = 75$ $\mu$m employed by Bilston et al.\cite{bilston2003arterial}, the resulting velocity increases to $u'_{\rm mean}=14.26$ $\mu$m/s. These velocities naturally compare with the average axial flow velocities on the order of 20$\mu$m/s observed experimentally in mouse pial periarterial spaces by Mestre et al.~\cite{mestre2018flow}, and inverse computational modelling estimates of the perivascular velocities sufficient to explain human solute transport observations of 2-14 $\mu$m/s~\cite{vinje2023human}.

In this context, we also note that Bojarskaite et al~\cite{bojarskaite2023sleep} demonstrate that the properties of the low frequency vascular dynamics change with sleep and locomotion, in particular the dynamics of pial arteries and penetrating arterioles with non-REM, intermediate and REM sleep in mice. Their study shows that the amplitude of the very low frequency (0.1--0.3 Hz) and low frequency (0.3--1.0 Hz) waves were higher (both of in terms of the vessel diameter and the PVS width) during non-REM and intermediate sleep states and during locomotion. Given the quadratic dependence on $\varepsilon$ exhibited in~\eqref{umeaneq}, it is clear that this increase in the wave amplitude produce a significant augmentation of the resulting flow. In addition, the blood vessel diameter ($2 r'_o$) was found to be larger and the PVS width ($r'_e - r'_o$) smaller during REM sleep compared to other sleep states and wakefulness~\cite{bojarskaite2023sleep}. As revealed by~\eqref{umeaneq}, the change in the cross section, resulting in an increased value of $r'_o$, a reduced value of $\beta$ and an associated increase in $\Delta$, enhance the induced motion. For instance, using in~\eqref{umeaneq} the values reported by Bojarskaite et al~\cite{bojarskaite2023sleep}, one can see that the increased median arteriole diameter during REM sleep ($\simeq 14 \mu$m i.e.~$r'_o = 7 \mu$m vs. $r'_o = 6 \mu$m) and decreased median PVS width ($\simeq 5 \mu$m, i.e~$r'_e = 12 \mu$m vs $r'_e = 13 \mu$m) result in an increase in net velocity by a factor of 3.2: $u'_{\rm mean} = 6.31 \mu$m/s compared to the baseline $u'_{\rm mean} = 1.96 \mu$m/s.

\begin{widetext}
\begin{figure}
  \begin{center}
  \begin{subfigure}{0.99\textwidth}
    \includegraphics[width=\textwidth]{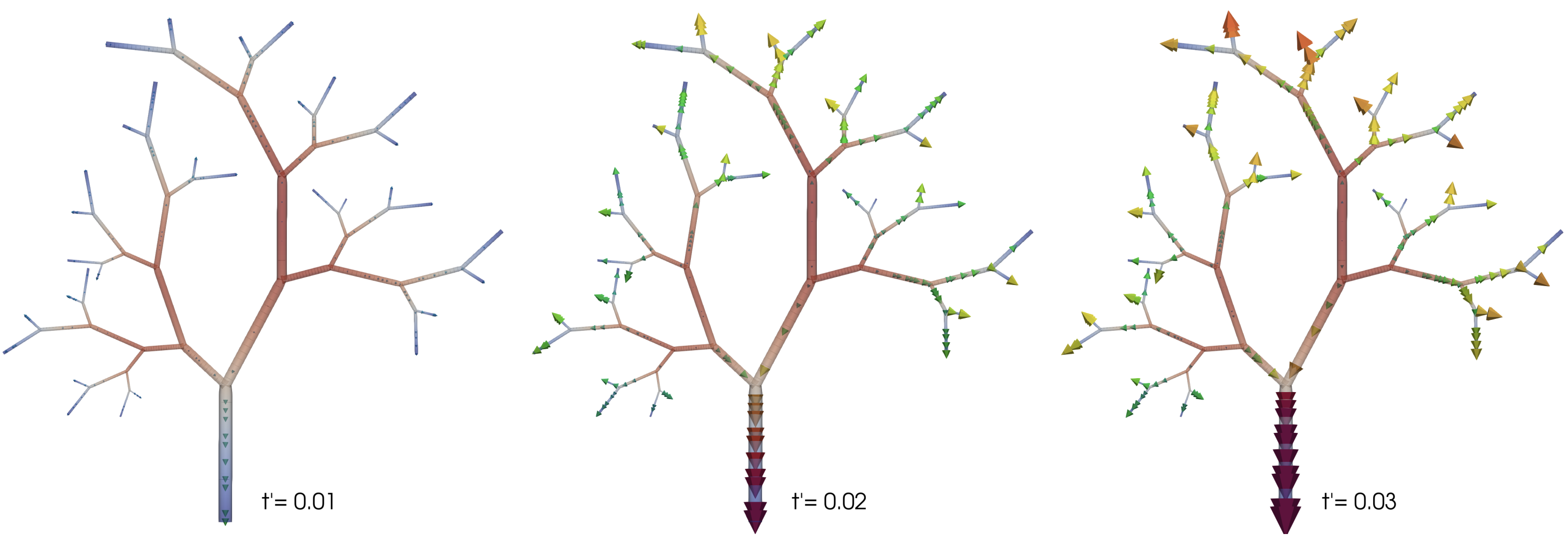}
    \caption{Snapshots of cardiac driven arterial expansion}
  \end{subfigure}
  \begin{subfigure}{0.99\textwidth}
    \includegraphics[width=\textwidth]{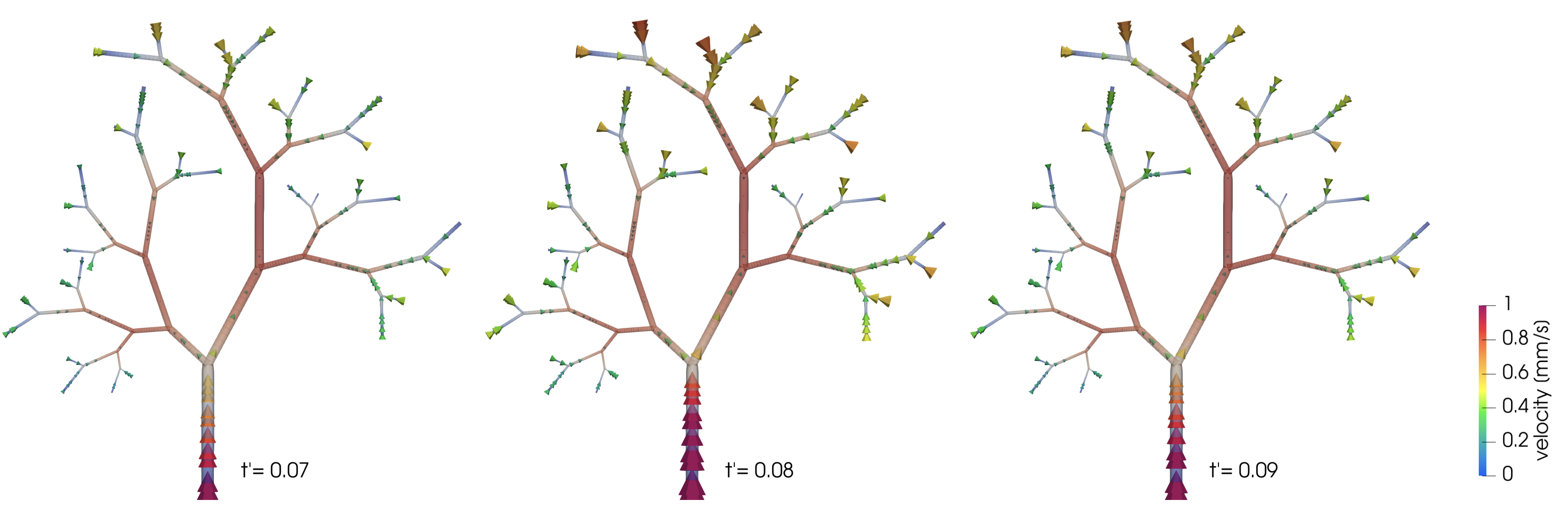}
    \caption{Snapshots of cardiac driven arterial contraction}
  \end{subfigure}
  \begin{subfigure}{0.99\textwidth}
    \includegraphics[width=\textwidth]{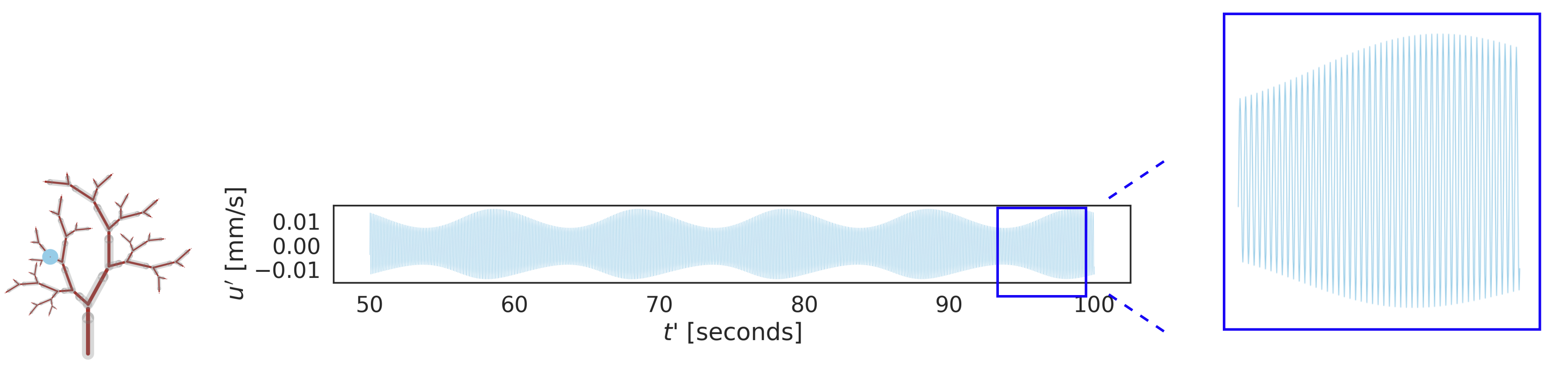}
    \caption{Interior node flow for five vasomotion cycles.}
  \end{subfigure} 
  \end{center}
  \caption{Pressure and velocity results for combined vasomotion and cardiac peristaltic pumping. Panels (a) and (b) show snapshots of the solution in time. In panel (a), cardiac driven arterial expansion occurs, resulting in the outward movement of fluid from the periarterial tree. Subsequently, in panel (b), cardiac driven arterial contraction causes the fluid to be drawn back into the arterial system. Panel (c) presents a plot of the axial velocity $u'$ at an interior node, highlighting how lower frequency vasomotion further modifies the flow rates over time. It is noteworthy that while the cardiac expansion contributes noticeably to the flow velocities (as evident in panel (c)), it does not have a significant impact on the net flow rates, as shown in Figure \ref{fig:interaction}.
  }
  \label{fig:snapshots}
\end{figure}    
\end{widetext}

\clearpage

\subsection{Simulation of cardiac and vasomotion in arterial tree}

Finally, we simulate net flow due to cardiac and vasomotion peristaltic pumping in a perivascular tree corresponding to the pial arterial network of a rat.  The arterial tree comprises six generations with decreasing radii constructed based on Murray's law, which describes the relationship between the radii of parent and daughter vessels. The initial branch of the tree is of radius $r_{0_0}'=1$ mm. Each branch has a length of $L_i=10r_{0_i}$. For each parent vessel $i$, the radii of the daughter vessels $j$ and $k$ are related by
\begin{align}
    r_{o_i}'^3 = r_{o_j}'^3 + r_{o_k}'^3,
\end{align}
and the ratio between the radii of the two daughter vessels is set to $r_{o_k}'/r_{o_j}'=0.8$.

Peristaltic pumping is modelled via periodic contraction and expansion of the inner vessel radius, 
\begin{align}
    r_a'/r_0' = 1 + \epsilon_c \sin(k_c x' - \omega_c t') + \epsilon_v \sin(k_v x' - \omega_v t') 
\end{align}
with wave numbers $k_c=2\pi /\lambda_c$, $k_v=2\pi /\lambda_v$ and angular frequency $\omega_c=2\pi f_c$ and $\omega_v=2\pi f_v$ defined as before. The travelling wave parameters are assigned using values from Table \ref{tab:physiological}. Specifically, $\epsilon_c=0.01$ denotes the amplitude of the rapid, small amplitude cardiac travelling waves, with a wavelength of $\lambda_c=100$ mm and a frequency of $f_c=10$ Hz. On the other hand, $\epsilon_v=0.1$ represents the amplitude of the slow, medium amplitude vasomotion waves, characterized by a wavelength of $\lambda_v=4$ mm and a frequency of $f_v=0.1$ Hz.

\begin{widetext}
\begin{figure}
  \begin{center}
  \includegraphics[width=0.9\textwidth]{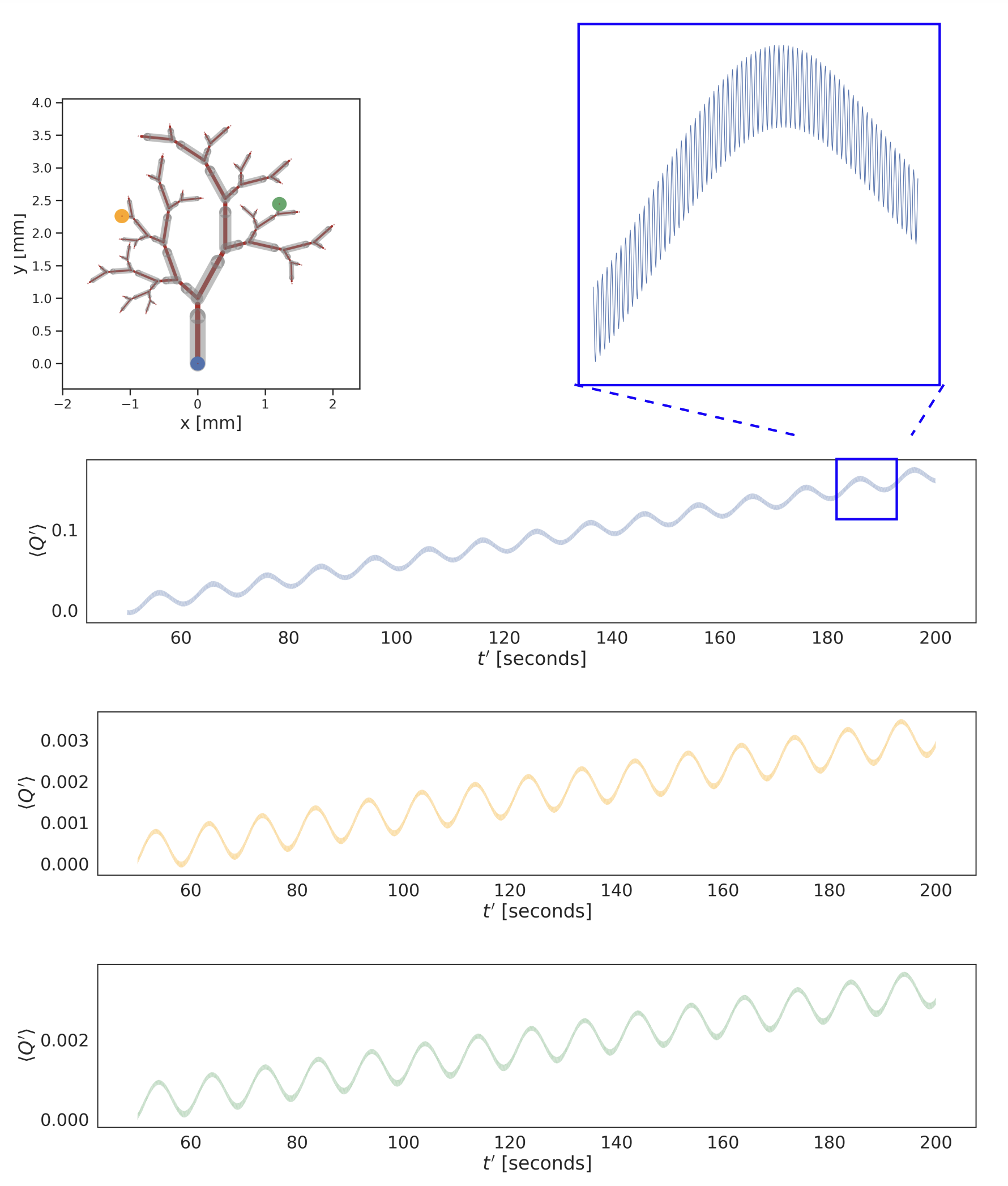}
  \end{center}
  \caption{Net flow over time driven by combined cardiac and vasomotion peristaltic pumping in a periarterial tree. (Top left) The periarterial tree is scaled to correspond to the pial arterial network of the rat cortex, with first vessel length $L=1$mm and radius $0.1$mm. Peristaltic pumping is modelled via travelling, sinusoidal contractions, with the cardiac component having small amplitude ($\epsilon_c=0.01$), wavelength $\lambda=100$mm and freq=10Hz. The vasomotion component is moderate amplitude ($\epsilon=0.1$) with shorter wavelength $\lambda=4$mm and lower frequency 0.1Hz. This results in a directional but oscillatory flow out of the root node (closeup shown in top right) and leaf nodes (middle and bottom plots), with the majority of the net flow caused by vasomotion peristaltic pumping.}
  \label{fig:interaction}
\end{figure}
\end{widetext}

Figure \ref{fig:snapshots} shows snapshots of the solutions over one cardiac cycle, as well as a plot of the flow velocity for five vasomotion cycles at an interior node in the network. Notably, cardiac and vasomotion contribute to velocities of the same order of magnitude; the high frequency of cardiac driven wall motion compensates for its comparatively small amplitude. Interestingly, the examination of net flow patterns over time reveals a different perspective. Figure \ref{fig:interaction} displays the temporal profiles of net flow at the inlet (depicted in blue) and two outlet nodes (represented by green and yellow). Despite the inherent oscillations in the flow, a clear directional trend emerges, as indicated by the continuous increase in net flow through both the inlet and outlet nodes. The net flow is clearly linked to the effect of vasomotion mediated peristaltic pumping.

\clearpage

\section{Concluding remarks}
\label{sec:conclusions}

The main findings and conclusions are as follows.
\begin{itemize}
\item 
  Our work presents a novel reduced-order model description for efficient computation of peristaltic flow in general perivascular networks parametrized by physiological parameters. The model assumes that (i) the peristaltic-wave amplitude is small and its wavelength is much larger than characteristic perivascular transverse size, (ii) the flow is dominated by viscous forces, and (iii) the peristaltic wave propagates without significant reflections at the arterial bifurcations. 
\item   
  The theoretical model demonstrates that vasomotion, at the wave speeds and frequencies reported in the literature and in contrast to e.g.~arterial pulsations induced by the cardiac pulse wave, can induce non-negligible net perivascular fluid flow at the order of some $\mu$m/s. 
\item   
  Moreover, increases in wave amplitudes or increased vascular diameters combined with decreased PVS widths, i.e.~modulations in the low-frequency vascular dynamics known to be associated with different stages of sleep, will enhance the net flow induced by peristalsis.  
\end{itemize}
Future extensions of the model should consider, in particular, effects of fluid loss into the surrounding tissue or vasculature via AQP4 channels in astrocyte endfeet, astrocyte endfeet gaps, or across the blood-brain-barrier. In addition, in improving predictive capabilities, the model should incorporate a more accurate description of the peristaltic wave, including reflections at arterial bifurcations, thereby enabling a precise quantification of the nonlinear interactions between the pressure fluctuations and the peristaltic wave.

\begin{acknowledgments}
MER graciously acknowledges discussions with Dr.~Cécile Daversin-Catty on the topic of modelling perivascular flow.

MER acknowledges support and funding from the Research Council of
Norway (RCN) via FRIPRO grant agreement \#324239 (EMIx), and the
U.S.-Norway Fulbright Foundation for Educational Exchange. The work of
ALS was supported by the National Institute of Neurological Disorders
and Stroke through contract no. 1R01NS120343-01.
\end{acknowledgments}

\section*{Data Availability Statement}
The software and scripts used to generate the graphics, data and results reported in this manuscript are openly and publicly available for inspection and download at~\href{https://github.com/scientificcomputing/perivascular-peristalsis}{https://github.com/scientificcomputing/perivascular-peristalsis}.

\bibliography{JAP_GSR_2023}

\providecommand{\noopsort}[1]{}\providecommand{\singleletter}[1]{#1}%
\begin{thebibliography}{45}%
\makeatletter
\providecommand \@ifxundefined [1]{%
 \@ifx{#1\undefined}
}%
\providecommand \@ifnum [1]{%
 \ifnum #1\expandafter \@firstoftwo
 \else \expandafter \@secondoftwo
 \fi
}%
\providecommand \@ifx [1]{%
 \ifx #1\expandafter \@firstoftwo
 \else \expandafter \@secondoftwo
 \fi
}%
\providecommand \natexlab [1]{#1}%
\providecommand \enquote  [1]{``#1''}%
\providecommand \bibnamefont  [1]{#1}%
\providecommand \bibfnamefont [1]{#1}%
\providecommand \citenamefont [1]{#1}%
\providecommand \href@noop [0]{\@secondoftwo}%
\providecommand \href [0]{\begingroup \@sanitize@url \@href}%
\providecommand \@href[1]{\@@startlink{#1}\@@href}%
\providecommand \@@href[1]{\endgroup#1\@@endlink}%
\providecommand \@sanitize@url [0]{\catcode `\\12\catcode `\$12\catcode
  `\&12\catcode `\#12\catcode `\^12\catcode `\_12\catcode `\%12\relax}%
\providecommand \@@startlink[1]{}%
\providecommand \@@endlink[0]{}%
\providecommand \url  [0]{\begingroup\@sanitize@url \@url }%
\providecommand \@url [1]{\endgroup\@href {#1}{\urlprefix }}%
\providecommand \urlprefix  [0]{URL }%
\providecommand \Eprint [0]{\href }%
\providecommand \doibase [0]{http://dx.doi.org/}%
\providecommand \selectlanguage [0]{\@gobble}%
\providecommand \bibinfo  [0]{\@secondoftwo}%
\providecommand \bibfield  [0]{\@secondoftwo}%
\providecommand \translation [1]{[#1]}%
\providecommand \BibitemOpen [0]{}%
\providecommand \bibitemStop [0]{}%
\providecommand \bibitemNoStop [0]{.\EOS\space}%
\providecommand \EOS [0]{\spacefactor3000\relax}%
\providecommand \BibitemShut  [1]{\csname bibitem#1\endcsname}%
\let\auto@bib@innerbib\@empty
\bibitem [{\citenamefont {Flexner}(1933)}]{flexner1933some}%
  \BibitemOpen
  \bibfield  {author} {\bibinfo {author} {\bibfnamefont {L.~B.}\ \bibnamefont
  {Flexner}},\ }\bibfield  {title} {\enquote {\bibinfo {title} {Some problems
  of the origin, circulation and absorption of the cerebrospinal fluid},}\
  }\href@noop {} {\bibfield  {journal} {\bibinfo  {journal} {The Quarterly
  Review of Biology}\ }\textbf {\bibinfo {volume} {8}},\ \bibinfo {pages}
  {397--422} (\bibinfo {year} {1933})}\BibitemShut {NoStop}%
\bibitem [{\citenamefont {Rennels}\ \emph {et~al.}(1985)\citenamefont
  {Rennels}, \citenamefont {Gregory}, \citenamefont {Blaumanis}, \citenamefont
  {Fujimoto},\ and\ \citenamefont {Grady}}]{rennels1985evidence}%
  \BibitemOpen
  \bibfield  {author} {\bibinfo {author} {\bibfnamefont {M.}~\bibnamefont
  {Rennels}}, \bibinfo {author} {\bibfnamefont {T.}~\bibnamefont {Gregory}},
  \bibinfo {author} {\bibfnamefont {O.}~\bibnamefont {Blaumanis}}, \bibinfo
  {author} {\bibfnamefont {K.}~\bibnamefont {Fujimoto}}, \ and\ \bibinfo
  {author} {\bibfnamefont {P.}~\bibnamefont {Grady}},\ }\bibfield  {title}
  {\enquote {\bibinfo {title} {Evidence for a ‘paravascular’ fluid
  circulation in the mammalian central nervous system, provided by the rapid
  distribution of tracer protein throughout the brain from the subarachnoid
  space},}\ }\href@noop {} {\bibfield  {journal} {\bibinfo  {journal} {Brain
  Res.}\ }\textbf {\bibinfo {volume} {326}},\ \bibinfo {pages} {47--63}
  (\bibinfo {year} {1985})}\BibitemShut {NoStop}%
\bibitem [{\citenamefont {Ichimura}, \citenamefont {Fraser},\ and\
  \citenamefont {Cserr}(1991)}]{ichimura1991distribution}%
  \BibitemOpen
  \bibfield  {author} {\bibinfo {author} {\bibfnamefont {T.}~\bibnamefont
  {Ichimura}}, \bibinfo {author} {\bibfnamefont {P.}~\bibnamefont {Fraser}}, \
  and\ \bibinfo {author} {\bibfnamefont {H.~F.}\ \bibnamefont {Cserr}},\
  }\bibfield  {title} {\enquote {\bibinfo {title} {Distribution of
  extracellular tracers in perivascular spaces of the rat brain},}\ }\href@noop
  {} {\bibfield  {journal} {\bibinfo  {journal} {Brain research}\ }\textbf
  {\bibinfo {volume} {545}},\ \bibinfo {pages} {103--113} (\bibinfo {year}
  {1991})}\BibitemShut {NoStop}%
\bibitem [{\citenamefont {Iliff}\ \emph {et~al.}(2012)\citenamefont {Iliff},
  \citenamefont {Wang}, \citenamefont {Liao}, \citenamefont {Plogg},
  \citenamefont {Peng}, \citenamefont {Gundersen}, \citenamefont {Benveniste},
  \citenamefont {Vates}, \citenamefont {Deane}, \citenamefont {Goldman} \emph
  {et~al.}}]{iliff2012paravascular}%
  \BibitemOpen
  \bibfield  {author} {\bibinfo {author} {\bibfnamefont {J.~J.}\ \bibnamefont
  {Iliff}}, \bibinfo {author} {\bibfnamefont {M.}~\bibnamefont {Wang}},
  \bibinfo {author} {\bibfnamefont {Y.}~\bibnamefont {Liao}}, \bibinfo {author}
  {\bibfnamefont {B.~A.}\ \bibnamefont {Plogg}}, \bibinfo {author}
  {\bibfnamefont {W.}~\bibnamefont {Peng}}, \bibinfo {author} {\bibfnamefont
  {G.~A.}\ \bibnamefont {Gundersen}}, \bibinfo {author} {\bibfnamefont
  {H.}~\bibnamefont {Benveniste}}, \bibinfo {author} {\bibfnamefont {G.~E.}\
  \bibnamefont {Vates}}, \bibinfo {author} {\bibfnamefont {R.}~\bibnamefont
  {Deane}}, \bibinfo {author} {\bibfnamefont {S.~A.}\ \bibnamefont {Goldman}},
  \emph {et~al.},\ }\bibfield  {title} {\enquote {\bibinfo {title} {A
  paravascular pathway facilitates {CSF} flow through the brain parenchyma and
  the clearance of interstitial solutes, including amyloid $\beta$},}\
  }\href@noop {} {\bibfield  {journal} {\bibinfo  {journal} {Science
  translational medicine}\ }\textbf {\bibinfo {volume} {4}},\ \bibinfo {pages}
  {147ra111--147ra111} (\bibinfo {year} {2012})}\BibitemShut {NoStop}%
\bibitem [{\citenamefont {Iliff}\ \emph {et~al.}(2013)\citenamefont {Iliff},
  \citenamefont {Wang}, \citenamefont {Zeppenfeld}, \citenamefont
  {Venkataraman}, \citenamefont {Plog}, \citenamefont {Liao}, \citenamefont
  {Deane},\ and\ \citenamefont {Nedergaard}}]{iliff2013cerebral}%
  \BibitemOpen
  \bibfield  {author} {\bibinfo {author} {\bibfnamefont {J.~J.}\ \bibnamefont
  {Iliff}}, \bibinfo {author} {\bibfnamefont {M.}~\bibnamefont {Wang}},
  \bibinfo {author} {\bibfnamefont {D.~M.}\ \bibnamefont {Zeppenfeld}},
  \bibinfo {author} {\bibfnamefont {A.}~\bibnamefont {Venkataraman}}, \bibinfo
  {author} {\bibfnamefont {B.~A.}\ \bibnamefont {Plog}}, \bibinfo {author}
  {\bibfnamefont {Y.}~\bibnamefont {Liao}}, \bibinfo {author} {\bibfnamefont
  {R.}~\bibnamefont {Deane}}, \ and\ \bibinfo {author} {\bibfnamefont
  {M.}~\bibnamefont {Nedergaard}},\ }\bibfield  {title} {\enquote {\bibinfo
  {title} {Cerebral arterial pulsation drives paravascular {CSF}--interstitial
  fluid exchange in the murine brain},}\ }\href@noop {} {\bibfield  {journal}
  {\bibinfo  {journal} {Journal of Neuroscience}\ }\textbf {\bibinfo {volume}
  {33}},\ \bibinfo {pages} {18190--18199} (\bibinfo {year} {2013})}\BibitemShut
  {NoStop}%
\bibitem [{\citenamefont {Ngo}, \citenamefont {Sarkaria},\ and\ \citenamefont
  {Harley}(2022)}]{ngo2022perivascular}%
  \BibitemOpen
  \bibfield  {author} {\bibinfo {author} {\bibfnamefont {M.~T.}\ \bibnamefont
  {Ngo}}, \bibinfo {author} {\bibfnamefont {J.~N.}\ \bibnamefont {Sarkaria}}, \
  and\ \bibinfo {author} {\bibfnamefont {B.~A.}\ \bibnamefont {Harley}},\
  }\bibfield  {title} {\enquote {\bibinfo {title} {Perivascular stromal cells
  instruct glioblastoma invasion, proliferation, and therapeutic response
  within an engineered brain perivascular niche model},}\ }\href@noop {}
  {\bibfield  {journal} {\bibinfo  {journal} {Advanced Science}\ }\textbf
  {\bibinfo {volume} {9}},\ \bibinfo {pages} {2201888} (\bibinfo {year}
  {2022})}\BibitemShut {NoStop}%
\bibitem [{\citenamefont {Lilius}\ \emph {et~al.}(2023)\citenamefont {Lilius},
  \citenamefont {Mortensen}, \citenamefont {Deville}, \citenamefont {Lohela},
  \citenamefont {St{\ae}ger}, \citenamefont {Sigurdsson}, \citenamefont
  {Fiordaliso}, \citenamefont {Rosenholm}, \citenamefont {Kamphuis},
  \citenamefont {Beekman} \emph {et~al.}}]{lilius2023glymphatic}%
  \BibitemOpen
  \bibfield  {author} {\bibinfo {author} {\bibfnamefont {T.~O.}\ \bibnamefont
  {Lilius}}, \bibinfo {author} {\bibfnamefont {K.~N.}\ \bibnamefont
  {Mortensen}}, \bibinfo {author} {\bibfnamefont {C.}~\bibnamefont {Deville}},
  \bibinfo {author} {\bibfnamefont {T.~J.}\ \bibnamefont {Lohela}}, \bibinfo
  {author} {\bibfnamefont {F.~F.}\ \bibnamefont {St{\ae}ger}}, \bibinfo
  {author} {\bibfnamefont {B.}~\bibnamefont {Sigurdsson}}, \bibinfo {author}
  {\bibfnamefont {E.~M.}\ \bibnamefont {Fiordaliso}}, \bibinfo {author}
  {\bibfnamefont {M.}~\bibnamefont {Rosenholm}}, \bibinfo {author}
  {\bibfnamefont {C.}~\bibnamefont {Kamphuis}}, \bibinfo {author}
  {\bibfnamefont {F.~J.}\ \bibnamefont {Beekman}},  \emph {et~al.},\ }\bibfield
   {title} {\enquote {\bibinfo {title} {Glymphatic-assisted perivascular brain
  delivery of intrathecal small gold nanoparticles},}\ }\href@noop {}
  {\bibfield  {journal} {\bibinfo  {journal} {Journal of Controlled Release}\
  }\textbf {\bibinfo {volume} {355}},\ \bibinfo {pages} {135--148} (\bibinfo
  {year} {2023})}\BibitemShut {NoStop}%
\bibitem [{\citenamefont {Nedergaard}\ and\ \citenamefont
  {Goldman}(2020)}]{nedergaard2020glymphatic}%
  \BibitemOpen
  \bibfield  {author} {\bibinfo {author} {\bibfnamefont {M.}~\bibnamefont
  {Nedergaard}}\ and\ \bibinfo {author} {\bibfnamefont {S.~A.}\ \bibnamefont
  {Goldman}},\ }\bibfield  {title} {\enquote {\bibinfo {title} {Glymphatic
  failure as a final common pathway to dementia},}\ }\href@noop {} {\bibfield
  {journal} {\bibinfo  {journal} {Science}\ }\textbf {\bibinfo {volume}
  {370}},\ \bibinfo {pages} {50--56} (\bibinfo {year} {2020})}\BibitemShut
  {NoStop}%
\bibitem [{\citenamefont {Weller}\ \emph {et~al.}(2008)\citenamefont {Weller},
  \citenamefont {Subash}, \citenamefont {Preston}, \citenamefont {Mazanti},\
  and\ \citenamefont {Carare}}]{weller2008perivascular}%
  \BibitemOpen
  \bibfield  {author} {\bibinfo {author} {\bibfnamefont {R.~O.}\ \bibnamefont
  {Weller}}, \bibinfo {author} {\bibfnamefont {M.}~\bibnamefont {Subash}},
  \bibinfo {author} {\bibfnamefont {S.~D.}\ \bibnamefont {Preston}}, \bibinfo
  {author} {\bibfnamefont {I.}~\bibnamefont {Mazanti}}, \ and\ \bibinfo
  {author} {\bibfnamefont {R.~O.}\ \bibnamefont {Carare}},\ }\bibfield  {title}
  {\enquote {\bibinfo {title} {Perivascular drainage of amyloid-beta peptides
  from the brain and its failure in cerebral amyloid angiopathy and alzheimer's
  disease.}}\ }\href@noop {} {\bibfield  {journal} {\bibinfo  {journal} {Brain
  pathology (Zurich, Switzerland)}\ }\textbf {\bibinfo {volume} {18}},\
  \bibinfo {pages} {253--266} (\bibinfo {year} {2008})}\BibitemShut {NoStop}%
\bibitem [{\citenamefont {Mestre}\ \emph {et~al.}(2022)\citenamefont {Mestre},
  \citenamefont {Verma}, \citenamefont {Greene}, \citenamefont {Lin},
  \citenamefont {Ladron-de Guevara}, \citenamefont {Sweeney}, \citenamefont
  {Liu}, \citenamefont {Thomas}, \citenamefont {Galloway}, \citenamefont
  {de~Mesy~Bentley} \emph {et~al.}}]{mestre2022periarteriolar}%
  \BibitemOpen
  \bibfield  {author} {\bibinfo {author} {\bibfnamefont {H.}~\bibnamefont
  {Mestre}}, \bibinfo {author} {\bibfnamefont {N.}~\bibnamefont {Verma}},
  \bibinfo {author} {\bibfnamefont {T.~D.}\ \bibnamefont {Greene}}, \bibinfo
  {author} {\bibfnamefont {L.~A.}\ \bibnamefont {Lin}}, \bibinfo {author}
  {\bibfnamefont {A.}~\bibnamefont {Ladron-de Guevara}}, \bibinfo {author}
  {\bibfnamefont {A.~M.}\ \bibnamefont {Sweeney}}, \bibinfo {author}
  {\bibfnamefont {G.}~\bibnamefont {Liu}}, \bibinfo {author} {\bibfnamefont
  {V.~K.}\ \bibnamefont {Thomas}}, \bibinfo {author} {\bibfnamefont {C.~A.}\
  \bibnamefont {Galloway}}, \bibinfo {author} {\bibfnamefont {K.~L.}\
  \bibnamefont {de~Mesy~Bentley}},  \emph {et~al.},\ }\bibfield  {title}
  {\enquote {\bibinfo {title} {Periarteriolar spaces modulate cerebrospinal
  fluid transport into brain and demonstrate altered morphology in aging and
  {Alzheimer}’s disease},}\ }\href@noop {} {\bibfield  {journal} {\bibinfo
  {journal} {Nature Communications}\ }\textbf {\bibinfo {volume} {13}},\
  \bibinfo {pages} {3897} (\bibinfo {year} {2022})}\BibitemShut {NoStop}%
\bibitem [{\citenamefont {Zhang}\ \emph {et~al.}(2023)\citenamefont {Zhang},
  \citenamefont {Zhang}, \citenamefont {He}, \citenamefont {Li}, \citenamefont
  {Meng}, \citenamefont {Mao}, \citenamefont {Li}, \citenamefont {Xue},
  \citenamefont {Gui}, \citenamefont {Zhang} \emph
  {et~al.}}]{zhang2023interaction}%
  \BibitemOpen
  \bibfield  {author} {\bibinfo {author} {\bibfnamefont {Y.}~\bibnamefont
  {Zhang}}, \bibinfo {author} {\bibfnamefont {C.}~\bibnamefont {Zhang}},
  \bibinfo {author} {\bibfnamefont {X.-Z.}\ \bibnamefont {He}}, \bibinfo
  {author} {\bibfnamefont {Z.-H.}\ \bibnamefont {Li}}, \bibinfo {author}
  {\bibfnamefont {J.-C.}\ \bibnamefont {Meng}}, \bibinfo {author}
  {\bibfnamefont {R.-T.}\ \bibnamefont {Mao}}, \bibinfo {author} {\bibfnamefont
  {X.}~\bibnamefont {Li}}, \bibinfo {author} {\bibfnamefont {R.}~\bibnamefont
  {Xue}}, \bibinfo {author} {\bibfnamefont {Q.}~\bibnamefont {Gui}}, \bibinfo
  {author} {\bibfnamefont {G.-X.}\ \bibnamefont {Zhang}},  \emph {et~al.},\
  }\bibfield  {title} {\enquote {\bibinfo {title} {Interaction between the
  glymphatic system and $\alpha$-synuclein in parkinson’s disease},}\
  }\href@noop {} {\bibfield  {journal} {\bibinfo  {journal} {Molecular
  Neurobiology}\ ,\ \bibinfo {pages} {1--14}} (\bibinfo {year}
  {2023})}\BibitemShut {NoStop}%
\bibitem [{\citenamefont {Rasmussen}, \citenamefont {Mestre},\ and\
  \citenamefont {Nedergaard}(2018)}]{rasmussen2018glymphatic}%
  \BibitemOpen
  \bibfield  {author} {\bibinfo {author} {\bibfnamefont {M.~K.}\ \bibnamefont
  {Rasmussen}}, \bibinfo {author} {\bibfnamefont {H.}~\bibnamefont {Mestre}}, \
  and\ \bibinfo {author} {\bibfnamefont {M.}~\bibnamefont {Nedergaard}},\
  }\bibfield  {title} {\enquote {\bibinfo {title} {The glymphatic pathway in
  neurological disorders},}\ }\href@noop {} {\bibfield  {journal} {\bibinfo
  {journal} {The Lancet Neurology}\ }\textbf {\bibinfo {volume} {17}},\
  \bibinfo {pages} {1016--1024} (\bibinfo {year} {2018})}\BibitemShut {NoStop}%
\bibitem [{\citenamefont {von Holstein-Rathlou}, \citenamefont {Petersen},\
  and\ \citenamefont {Nedergaard}(2018)}]{von2018voluntary}%
  \BibitemOpen
  \bibfield  {author} {\bibinfo {author} {\bibfnamefont {S.}~\bibnamefont {von
  Holstein-Rathlou}}, \bibinfo {author} {\bibfnamefont {N.~C.}\ \bibnamefont
  {Petersen}}, \ and\ \bibinfo {author} {\bibfnamefont {M.}~\bibnamefont
  {Nedergaard}},\ }\bibfield  {title} {\enquote {\bibinfo {title} {Voluntary
  running enhances glymphatic influx in awake behaving, young mice},}\
  }\href@noop {} {\bibfield  {journal} {\bibinfo  {journal} {Neuroscience
  letters}\ }\textbf {\bibinfo {volume} {662}},\ \bibinfo {pages} {253--258}
  (\bibinfo {year} {2018})}\BibitemShut {NoStop}%
\bibitem [{\citenamefont {Xie}\ \emph {et~al.}(2013)\citenamefont {Xie},
  \citenamefont {Kang}, \citenamefont {Xu}, \citenamefont {Chen}, \citenamefont
  {Liao}, \citenamefont {Thiyagarajan}, \citenamefont {O’Donnell},
  \citenamefont {Christensen}, \citenamefont {Nicholson}, \citenamefont {Iliff}
  \emph {et~al.}}]{xie2013sleep}%
  \BibitemOpen
  \bibfield  {author} {\bibinfo {author} {\bibfnamefont {L.}~\bibnamefont
  {Xie}}, \bibinfo {author} {\bibfnamefont {H.}~\bibnamefont {Kang}}, \bibinfo
  {author} {\bibfnamefont {Q.}~\bibnamefont {Xu}}, \bibinfo {author}
  {\bibfnamefont {M.~J.}\ \bibnamefont {Chen}}, \bibinfo {author}
  {\bibfnamefont {Y.}~\bibnamefont {Liao}}, \bibinfo {author} {\bibfnamefont
  {M.}~\bibnamefont {Thiyagarajan}}, \bibinfo {author} {\bibfnamefont
  {J.}~\bibnamefont {O’Donnell}}, \bibinfo {author} {\bibfnamefont {D.~J.}\
  \bibnamefont {Christensen}}, \bibinfo {author} {\bibfnamefont
  {C.}~\bibnamefont {Nicholson}}, \bibinfo {author} {\bibfnamefont {J.~J.}\
  \bibnamefont {Iliff}},  \emph {et~al.},\ }\bibfield  {title} {\enquote
  {\bibinfo {title} {Sleep drives metabolite clearance from the adult brain},}\
  }\href@noop {} {\bibfield  {journal} {\bibinfo  {journal} {science}\ }\textbf
  {\bibinfo {volume} {342}},\ \bibinfo {pages} {373--377} (\bibinfo {year}
  {2013})}\BibitemShut {NoStop}%
\bibitem [{\citenamefont {Bojarskaite}\ \emph {et~al.}(2023)\citenamefont
  {Bojarskaite}, \citenamefont {Vallet}, \citenamefont {Bj{\o}rnstad},
  \citenamefont {Gullestad~Binder}, \citenamefont {Cunen}, \citenamefont
  {Heuser}, \citenamefont {Kuchta}, \citenamefont {Mardal},\ and\ \citenamefont
  {Enger}}]{bojarskaite2023sleep}%
  \BibitemOpen
  \bibfield  {author} {\bibinfo {author} {\bibfnamefont {L.}~\bibnamefont
  {Bojarskaite}}, \bibinfo {author} {\bibfnamefont {A.}~\bibnamefont {Vallet}},
  \bibinfo {author} {\bibfnamefont {D.~M.}\ \bibnamefont {Bj{\o}rnstad}},
  \bibinfo {author} {\bibfnamefont {K.~M.}\ \bibnamefont {Gullestad~Binder}},
  \bibinfo {author} {\bibfnamefont {C.}~\bibnamefont {Cunen}}, \bibinfo
  {author} {\bibfnamefont {K.}~\bibnamefont {Heuser}}, \bibinfo {author}
  {\bibfnamefont {M.}~\bibnamefont {Kuchta}}, \bibinfo {author} {\bibfnamefont
  {K.-A.}\ \bibnamefont {Mardal}}, \ and\ \bibinfo {author} {\bibfnamefont
  {R.}~\bibnamefont {Enger}},\ }\bibfield  {title} {\enquote {\bibinfo {title}
  {Sleep cycle-dependent vascular dynamics in male mice and the predicted
  effects on perivascular cerebrospinal fluid flow and solute transport},}\
  }\href@noop {} {\bibfield  {journal} {\bibinfo  {journal} {Nature
  Communications}\ }\textbf {\bibinfo {volume} {14}},\ \bibinfo {pages} {953}
  (\bibinfo {year} {2023})}\BibitemShut {NoStop}%
\bibitem [{\citenamefont {Zhang}, \citenamefont {Inman},\ and\ \citenamefont
  {Weller}(1990)}]{zhang1990interrelationships}%
  \BibitemOpen
  \bibfield  {author} {\bibinfo {author} {\bibfnamefont {E.}~\bibnamefont
  {Zhang}}, \bibinfo {author} {\bibfnamefont {C.}~\bibnamefont {Inman}}, \ and\
  \bibinfo {author} {\bibfnamefont {R.}~\bibnamefont {Weller}},\ }\bibfield
  {title} {\enquote {\bibinfo {title} {{Interrelationships of the pia mater and
  the perivascular (Virchow-Robin) spaces in the human cerebrum}},}\
  }\href@noop {} {\bibfield  {journal} {\bibinfo  {journal} {Journal of
  anatomy}\ }\textbf {\bibinfo {volume} {170}},\ \bibinfo {pages} {111}
  (\bibinfo {year} {1990})}\BibitemShut {NoStop}%
\bibitem [{\citenamefont {Bedussi}\ \emph {et~al.}(2018)\citenamefont
  {Bedussi}, \citenamefont {Almasian}, \citenamefont {de~Vos}, \citenamefont
  {VanBavel},\ and\ \citenamefont {Bakker}}]{bedussi2018paravascular}%
  \BibitemOpen
  \bibfield  {author} {\bibinfo {author} {\bibfnamefont {B.}~\bibnamefont
  {Bedussi}}, \bibinfo {author} {\bibfnamefont {M.}~\bibnamefont {Almasian}},
  \bibinfo {author} {\bibfnamefont {J.}~\bibnamefont {de~Vos}}, \bibinfo
  {author} {\bibfnamefont {E.}~\bibnamefont {VanBavel}}, \ and\ \bibinfo
  {author} {\bibfnamefont {E.~N.}\ \bibnamefont {Bakker}},\ }\bibfield  {title}
  {\enquote {\bibinfo {title} {Paravascular spaces at the brain surface: Low
  resistance pathways for cerebrospinal fluid flow},}\ }\href@noop {}
  {\bibfield  {journal} {\bibinfo  {journal} {Journal of Cerebral Blood Flow \&
  Metabolism}\ }\textbf {\bibinfo {volume} {38}},\ \bibinfo {pages} {719--726}
  (\bibinfo {year} {2018})}\BibitemShut {NoStop}%
\bibitem [{\citenamefont {Wardlaw}\ \emph {et~al.}(2020)\citenamefont
  {Wardlaw}, \citenamefont {Benveniste}, \citenamefont {Nedergaard},
  \citenamefont {Zlokovic}, \citenamefont {Mestre}, \citenamefont {Lee},
  \citenamefont {Doubal}, \citenamefont {Brown}, \citenamefont {Ramirez},
  \citenamefont {MacIntosh} \emph {et~al.}}]{wardlaw2020perivascular}%
  \BibitemOpen
  \bibfield  {author} {\bibinfo {author} {\bibfnamefont {J.~M.}\ \bibnamefont
  {Wardlaw}}, \bibinfo {author} {\bibfnamefont {H.}~\bibnamefont {Benveniste}},
  \bibinfo {author} {\bibfnamefont {M.}~\bibnamefont {Nedergaard}}, \bibinfo
  {author} {\bibfnamefont {B.~V.}\ \bibnamefont {Zlokovic}}, \bibinfo {author}
  {\bibfnamefont {H.}~\bibnamefont {Mestre}}, \bibinfo {author} {\bibfnamefont
  {H.}~\bibnamefont {Lee}}, \bibinfo {author} {\bibfnamefont {F.~N.}\
  \bibnamefont {Doubal}}, \bibinfo {author} {\bibfnamefont {R.}~\bibnamefont
  {Brown}}, \bibinfo {author} {\bibfnamefont {J.}~\bibnamefont {Ramirez}},
  \bibinfo {author} {\bibfnamefont {B.~J.}\ \bibnamefont {MacIntosh}},  \emph
  {et~al.},\ }\bibfield  {title} {\enquote {\bibinfo {title} {Perivascular
  spaces in the brain: anatomy, physiology and pathology},}\ }\href@noop {}
  {\bibfield  {journal} {\bibinfo  {journal} {Nature Reviews Neurology}\
  }\textbf {\bibinfo {volume} {16}},\ \bibinfo {pages} {137--153} (\bibinfo
  {year} {2020})}\BibitemShut {NoStop}%
\bibitem [{\citenamefont {Tithof}\ \emph {et~al.}(2019)\citenamefont {Tithof},
  \citenamefont {Kelley}, \citenamefont {Mestre}, \citenamefont {Nedergaard},\
  and\ \citenamefont {Thomas}}]{tithof2019hydraulic}%
  \BibitemOpen
  \bibfield  {author} {\bibinfo {author} {\bibfnamefont {J.}~\bibnamefont
  {Tithof}}, \bibinfo {author} {\bibfnamefont {D.}~\bibnamefont {Kelley}},
  \bibinfo {author} {\bibfnamefont {H.}~\bibnamefont {Mestre}}, \bibinfo
  {author} {\bibfnamefont {M.}~\bibnamefont {Nedergaard}}, \ and\ \bibinfo
  {author} {\bibfnamefont {J.}~\bibnamefont {Thomas}},\ }\bibfield  {title}
  {\enquote {\bibinfo {title} {Hydraulic resistance of periarterial spaces in
  the brain},}\ }\href@noop {} {\bibfield  {journal} {\bibinfo  {journal}
  {Fluids Barriers CNS}\ }\textbf {\bibinfo {volume} {16}},\ \bibinfo {pages}
  {1--13} (\bibinfo {year} {2019})}\BibitemShut {NoStop}%
\bibitem [{\citenamefont {Raicevic}\ \emph {et~al.}(2023)\citenamefont
  {Raicevic}, \citenamefont {Forer}, \citenamefont {Ladron-de Guevara},
  \citenamefont {Nedergaard}, \citenamefont {Kelley}, \citenamefont {Boster}
  \emph {et~al.}}]{raicevic2023sizes}%
  \BibitemOpen
  \bibfield  {author} {\bibinfo {author} {\bibfnamefont {N.}~\bibnamefont
  {Raicevic}}, \bibinfo {author} {\bibfnamefont {J.}~\bibnamefont {Forer}},
  \bibinfo {author} {\bibfnamefont {A.}~\bibnamefont {Ladron-de Guevara}},
  \bibinfo {author} {\bibfnamefont {M.}~\bibnamefont {Nedergaard}}, \bibinfo
  {author} {\bibfnamefont {D.}~\bibnamefont {Kelley}}, \bibinfo {author}
  {\bibfnamefont {K.}~\bibnamefont {Boster}},  \emph {et~al.},\ }\bibfield
  {title} {\enquote {\bibinfo {title} {Sizes and shapes of perivascular spaces
  surrounding murine pial arteries.}}\ }\href@noop {} {\bibfield  {journal}
  {\bibinfo  {journal} {Research Square}\ ,\ \bibinfo {pages} {rs--3}}
  (\bibinfo {year} {2023})}\BibitemShut {NoStop}%
\bibitem [{\citenamefont {Bilston}\ \emph {et~al.}(2003)\citenamefont
  {Bilston}, \citenamefont {Fletcher}, \citenamefont {Brodbelt},\ and\
  \citenamefont {Stoodley}}]{bilston2003arterial}%
  \BibitemOpen
  \bibfield  {author} {\bibinfo {author} {\bibfnamefont {L.}~\bibnamefont
  {Bilston}}, \bibinfo {author} {\bibfnamefont {D.}~\bibnamefont {Fletcher}},
  \bibinfo {author} {\bibfnamefont {A.}~\bibnamefont {Brodbelt}}, \ and\
  \bibinfo {author} {\bibfnamefont {M.}~\bibnamefont {Stoodley}},\ }\bibfield
  {title} {\enquote {\bibinfo {title} {Arterial pulsation-driven cerebrospinal
  fluid flow in the perivascular space: a computational model},}\ }\href@noop
  {} {\bibfield  {journal} {\bibinfo  {journal} {Comput. Methods Biomech.
  Biomed. Engin.}\ }\textbf {\bibinfo {volume} {6}},\ \bibinfo {pages}
  {235--241} (\bibinfo {year} {2003})}\BibitemShut {NoStop}%
\bibitem [{\citenamefont {Vinje}, \citenamefont {Bakker},\ and\ \citenamefont
  {Rognes}(2021)}]{vinje2021brain}%
  \BibitemOpen
  \bibfield  {author} {\bibinfo {author} {\bibfnamefont {V.}~\bibnamefont
  {Vinje}}, \bibinfo {author} {\bibfnamefont {E.~N.}\ \bibnamefont {Bakker}}, \
  and\ \bibinfo {author} {\bibfnamefont {M.~E.}\ \bibnamefont {Rognes}},\
  }\bibfield  {title} {\enquote {\bibinfo {title} {Brain solute transport is
  more rapid in periarterial than perivenous spaces},}\ }\href@noop {}
  {\bibfield  {journal} {\bibinfo  {journal} {Scientific Reports}\ }\textbf
  {\bibinfo {volume} {11}},\ \bibinfo {pages} {1--11} (\bibinfo {year}
  {2021})}\BibitemShut {NoStop}%
\bibitem [{\citenamefont {Carr}\ \emph {et~al.}(2021)\citenamefont {Carr},
  \citenamefont {Thomas}, \citenamefont {Liu},\ and\ \citenamefont
  {Shang}}]{carr2021peristaltic}%
  \BibitemOpen
  \bibfield  {author} {\bibinfo {author} {\bibfnamefont {J.}~\bibnamefont
  {Carr}}, \bibinfo {author} {\bibfnamefont {J.}~\bibnamefont {Thomas}},
  \bibinfo {author} {\bibfnamefont {J.}~\bibnamefont {Liu}}, \ and\ \bibinfo
  {author} {\bibfnamefont {J.}~\bibnamefont {Shang}},\ }\bibfield  {title}
  {\enquote {\bibinfo {title} {Peristaltic pumping in thin non-axisymmetric
  annular tubes},}\ }\href@noop {} {\bibfield  {journal} {\bibinfo  {journal}
  {J. Fluid Mech.}\ }\textbf {\bibinfo {volume} {917}},\ \bibinfo {pages} {A10}
  (\bibinfo {year} {2021})}\BibitemShut {NoStop}%
\bibitem [{\citenamefont {Mestre}\ \emph {et~al.}(2018)\citenamefont {Mestre},
  \citenamefont {Tithof}, \citenamefont {Du}, \citenamefont {Song},
  \citenamefont {Peng}, \citenamefont {Sweeney}, \citenamefont {Olveda},
  \citenamefont {Thomas}, \citenamefont {Nedergaard},\ and\ \citenamefont
  {Kelley}}]{mestre2018flow}%
  \BibitemOpen
  \bibfield  {author} {\bibinfo {author} {\bibfnamefont {H.}~\bibnamefont
  {Mestre}}, \bibinfo {author} {\bibfnamefont {J.}~\bibnamefont {Tithof}},
  \bibinfo {author} {\bibfnamefont {T.}~\bibnamefont {Du}}, \bibinfo {author}
  {\bibfnamefont {W.}~\bibnamefont {Song}}, \bibinfo {author} {\bibfnamefont
  {W.}~\bibnamefont {Peng}}, \bibinfo {author} {\bibfnamefont {A.}~\bibnamefont
  {Sweeney}}, \bibinfo {author} {\bibfnamefont {G.}~\bibnamefont {Olveda}},
  \bibinfo {author} {\bibfnamefont {J.}~\bibnamefont {Thomas}}, \bibinfo
  {author} {\bibfnamefont {M.}~\bibnamefont {Nedergaard}}, \ and\ \bibinfo
  {author} {\bibfnamefont {D.}~\bibnamefont {Kelley}},\ }\bibfield  {title}
  {\enquote {\bibinfo {title} {Flow of cerebrospinal fluid is driven by
  arterial pulsations and is reduced in hypertension},}\ }\href@noop {}
  {\bibfield  {journal} {\bibinfo  {journal} {Nature Comm.}\ }\textbf {\bibinfo
  {volume} {9}},\ \bibinfo {pages} {1--9} (\bibinfo {year} {2018})}\BibitemShut
  {NoStop}%
\bibitem [{\citenamefont {Daversin-Catty}\ \emph {et~al.}(2020)\citenamefont
  {Daversin-Catty}, \citenamefont {Vinje}, \citenamefont {Mardal},\ and\
  \citenamefont {Rognes}}]{daversin2020mechanisms}%
  \BibitemOpen
  \bibfield  {author} {\bibinfo {author} {\bibfnamefont {C.}~\bibnamefont
  {Daversin-Catty}}, \bibinfo {author} {\bibfnamefont {V.}~\bibnamefont
  {Vinje}}, \bibinfo {author} {\bibfnamefont {K.-A.}\ \bibnamefont {Mardal}}, \
  and\ \bibinfo {author} {\bibfnamefont {M.~E.}\ \bibnamefont {Rognes}},\
  }\bibfield  {title} {\enquote {\bibinfo {title} {The mechanisms behind
  perivascular fluid flow},}\ }\href@noop {} {\bibfield  {journal} {\bibinfo
  {journal} {bioRxiv}\ } (\bibinfo {year} {2020})}\BibitemShut {NoStop}%
\bibitem [{\citenamefont {Kelley}\ \emph {et~al.}(2022)\citenamefont {Kelley},
  \citenamefont {Bohr}, \citenamefont {Hjorth}, \citenamefont {Holst},
  \citenamefont {Hrab{\v{e}}tov{\'a}}, \citenamefont {Kiviniemi}, \citenamefont
  {Lilius}, \citenamefont {Lundgaard}, \citenamefont {Mardal}, \citenamefont
  {Martens} \emph {et~al.}}]{kelley2022glymphatic}%
  \BibitemOpen
  \bibfield  {author} {\bibinfo {author} {\bibfnamefont {D.~H.}\ \bibnamefont
  {Kelley}}, \bibinfo {author} {\bibfnamefont {T.}~\bibnamefont {Bohr}},
  \bibinfo {author} {\bibfnamefont {P.~G.}\ \bibnamefont {Hjorth}}, \bibinfo
  {author} {\bibfnamefont {S.~C.}\ \bibnamefont {Holst}}, \bibinfo {author}
  {\bibfnamefont {S.}~\bibnamefont {Hrab{\v{e}}tov{\'a}}}, \bibinfo {author}
  {\bibfnamefont {V.}~\bibnamefont {Kiviniemi}}, \bibinfo {author}
  {\bibfnamefont {T.}~\bibnamefont {Lilius}}, \bibinfo {author} {\bibfnamefont
  {I.}~\bibnamefont {Lundgaard}}, \bibinfo {author} {\bibfnamefont {K.-A.}\
  \bibnamefont {Mardal}}, \bibinfo {author} {\bibfnamefont {E.~A.}\
  \bibnamefont {Martens}},  \emph {et~al.},\ }\bibfield  {title} {\enquote
  {\bibinfo {title} {The glymphatic system: Current understanding and
  modeling},}\ }\href@noop {} {\bibfield  {journal} {\bibinfo  {journal}
  {Iscience}\ ,\ \bibinfo {pages} {104987}} (\bibinfo {year}
  {2022})}\BibitemShut {NoStop}%
\bibitem [{\citenamefont {Vinje}\ \emph {et~al.}(2023)\citenamefont {Vinje},
  \citenamefont {Zapf}, \citenamefont {Ringstad}, \citenamefont {Eide},
  \citenamefont {Rognes},\ and\ \citenamefont {Mardal}}]{vinje2023human}%
  \BibitemOpen
  \bibfield  {author} {\bibinfo {author} {\bibfnamefont {V.}~\bibnamefont
  {Vinje}}, \bibinfo {author} {\bibfnamefont {B.}~\bibnamefont {Zapf}},
  \bibinfo {author} {\bibfnamefont {G.}~\bibnamefont {Ringstad}}, \bibinfo
  {author} {\bibfnamefont {P.~K.}\ \bibnamefont {Eide}}, \bibinfo {author}
  {\bibfnamefont {M.~E.}\ \bibnamefont {Rognes}}, \ and\ \bibinfo {author}
  {\bibfnamefont {K.-A.}\ \bibnamefont {Mardal}},\ }\bibfield  {title}
  {\enquote {\bibinfo {title} {Human brain solute transport quantified by
  glymphatic {MRI}-informed biophysics during sleep and sleep deprivation},}\
  }\href@noop {} {\bibfield  {journal} {\bibinfo  {journal} {bioRxiv}\ ,\
  \bibinfo {pages} {2023--01}} (\bibinfo {year} {2023})}\BibitemShut {NoStop}%
\bibitem [{\citenamefont {Asgari}, \citenamefont {De~Z{\'e}licourt},\ and\
  \citenamefont {Kurtcuoglu}(2016)}]{asgari2016glymphatic}%
  \BibitemOpen
  \bibfield  {author} {\bibinfo {author} {\bibfnamefont {M.}~\bibnamefont
  {Asgari}}, \bibinfo {author} {\bibfnamefont {D.}~\bibnamefont
  {De~Z{\'e}licourt}}, \ and\ \bibinfo {author} {\bibfnamefont
  {V.}~\bibnamefont {Kurtcuoglu}},\ }\bibfield  {title} {\enquote {\bibinfo
  {title} {Glymphatic solute transport does not require bulk flow},}\
  }\href@noop {} {\bibfield  {journal} {\bibinfo  {journal} {Scientific
  reports}\ }\textbf {\bibinfo {volume} {6}},\ \bibinfo {pages} {1--11}
  (\bibinfo {year} {2016})}\BibitemShut {NoStop}%
\bibitem [{\citenamefont {Martinac}\ and\ \citenamefont
  {Bilston}(2020)}]{martinac2020computational}%
  \BibitemOpen
  \bibfield  {author} {\bibinfo {author} {\bibfnamefont {A.~D.}\ \bibnamefont
  {Martinac}}\ and\ \bibinfo {author} {\bibfnamefont {L.~E.}\ \bibnamefont
  {Bilston}},\ }\bibfield  {title} {\enquote {\bibinfo {title} {Computational
  modelling of fluid and solute transport in the brain},}\ }\href@noop {}
  {\bibfield  {journal} {\bibinfo  {journal} {Biomechanics and modeling in
  mechanobiology}\ }\textbf {\bibinfo {volume} {19}},\ \bibinfo {pages}
  {781--800} (\bibinfo {year} {2020})}\BibitemShut {NoStop}%
\bibitem [{\citenamefont {Kedarasetti}, \citenamefont {Drew},\ and\
  \citenamefont {Costanzo}(2020)}]{kedarasetti2020arterial}%
  \BibitemOpen
  \bibfield  {author} {\bibinfo {author} {\bibfnamefont {R.~T.}\ \bibnamefont
  {Kedarasetti}}, \bibinfo {author} {\bibfnamefont {P.~J.}\ \bibnamefont
  {Drew}}, \ and\ \bibinfo {author} {\bibfnamefont {F.}~\bibnamefont
  {Costanzo}},\ }\bibfield  {title} {\enquote {\bibinfo {title} {Arterial
  pulsations drive oscillatory flow of {CSF} but not directional pumping},}\
  }\href@noop {} {\bibfield  {journal} {\bibinfo  {journal} {Scientific
  reports}\ }\textbf {\bibinfo {volume} {10}},\ \bibinfo {pages} {10102}
  (\bibinfo {year} {2020})}\BibitemShut {NoStop}%
\bibitem [{\citenamefont {van Veluw}\ \emph {et~al.}(2020)\citenamefont {van
  Veluw}, \citenamefont {Hou}, \citenamefont {Calvo-Rodriguez}, \citenamefont
  {Arbel-Ornath}, \citenamefont {Snyder}, \citenamefont {Frosch}, \citenamefont
  {Greenberg},\ and\ \citenamefont {Bacskai}}]{van2020vasomotion}%
  \BibitemOpen
  \bibfield  {author} {\bibinfo {author} {\bibfnamefont {S.~J.}\ \bibnamefont
  {van Veluw}}, \bibinfo {author} {\bibfnamefont {S.~S.}\ \bibnamefont {Hou}},
  \bibinfo {author} {\bibfnamefont {M.}~\bibnamefont {Calvo-Rodriguez}},
  \bibinfo {author} {\bibfnamefont {M.}~\bibnamefont {Arbel-Ornath}}, \bibinfo
  {author} {\bibfnamefont {A.~C.}\ \bibnamefont {Snyder}}, \bibinfo {author}
  {\bibfnamefont {M.~P.}\ \bibnamefont {Frosch}}, \bibinfo {author}
  {\bibfnamefont {S.~M.}\ \bibnamefont {Greenberg}}, \ and\ \bibinfo {author}
  {\bibfnamefont {B.~J.}\ \bibnamefont {Bacskai}},\ }\bibfield  {title}
  {\enquote {\bibinfo {title} {Vasomotion as a driving force for paravascular
  clearance in the awake mouse brain},}\ }\href@noop {} {\bibfield  {journal}
  {\bibinfo  {journal} {Neuron}\ }\textbf {\bibinfo {volume} {105}},\ \bibinfo
  {pages} {549--561} (\bibinfo {year} {2020})}\BibitemShut {NoStop}%
\bibitem [{\citenamefont {Munting}\ \emph {et~al.}(2023)\citenamefont
  {Munting}, \citenamefont {Bonnar}, \citenamefont {Kozberg}, \citenamefont
  {Auger}, \citenamefont {Hirschler}, \citenamefont {Hou}, \citenamefont
  {Greenberg}, \citenamefont {Bacskai},\ and\ \citenamefont {van
  Veluw}}]{munting2023spontaneous}%
  \BibitemOpen
  \bibfield  {author} {\bibinfo {author} {\bibfnamefont {L.~P.}\ \bibnamefont
  {Munting}}, \bibinfo {author} {\bibfnamefont {O.}~\bibnamefont {Bonnar}},
  \bibinfo {author} {\bibfnamefont {M.~G.}\ \bibnamefont {Kozberg}}, \bibinfo
  {author} {\bibfnamefont {C.~A.}\ \bibnamefont {Auger}}, \bibinfo {author}
  {\bibfnamefont {L.}~\bibnamefont {Hirschler}}, \bibinfo {author}
  {\bibfnamefont {S.~S.}\ \bibnamefont {Hou}}, \bibinfo {author} {\bibfnamefont
  {S.~M.}\ \bibnamefont {Greenberg}}, \bibinfo {author} {\bibfnamefont {B.~J.}\
  \bibnamefont {Bacskai}}, \ and\ \bibinfo {author} {\bibfnamefont {S.~J.}\
  \bibnamefont {van Veluw}},\ }\bibfield  {title} {\enquote {\bibinfo {title}
  {Spontaneous vasomotion propagates along pial arterioles in the awake mouse
  brain like stimulus-evoked vascular reactivity},}\ }\href@noop {} {\bibfield
  {journal} {\bibinfo  {journal} {Journal of Cerebral Blood Flow \&
  Metabolism}\ ,\ \bibinfo {pages} {0271678X231152550}} (\bibinfo {year}
  {2023})}\BibitemShut {NoStop}%
\bibitem [{\citenamefont {Ma}\ \emph {et~al.}(2019)\citenamefont {Ma},
  \citenamefont {Ries}, \citenamefont {Decker}, \citenamefont {M{\"u}ller},
  \citenamefont {Riner}, \citenamefont {B{\"u}cker}, \citenamefont
  {Fassbender}, \citenamefont {Detmar},\ and\ \citenamefont
  {Proulx}}]{ma2019rapid}%
  \BibitemOpen
  \bibfield  {author} {\bibinfo {author} {\bibfnamefont {Q.}~\bibnamefont
  {Ma}}, \bibinfo {author} {\bibfnamefont {M.}~\bibnamefont {Ries}}, \bibinfo
  {author} {\bibfnamefont {Y.}~\bibnamefont {Decker}}, \bibinfo {author}
  {\bibfnamefont {A.}~\bibnamefont {M{\"u}ller}}, \bibinfo {author}
  {\bibfnamefont {C.}~\bibnamefont {Riner}}, \bibinfo {author} {\bibfnamefont
  {A.}~\bibnamefont {B{\"u}cker}}, \bibinfo {author} {\bibfnamefont
  {K.}~\bibnamefont {Fassbender}}, \bibinfo {author} {\bibfnamefont
  {M.}~\bibnamefont {Detmar}}, \ and\ \bibinfo {author} {\bibfnamefont {S.~T.}\
  \bibnamefont {Proulx}},\ }\bibfield  {title} {\enquote {\bibinfo {title}
  {Rapid lymphatic efflux limits cerebrospinal fluid flow to the brain},}\
  }\href@noop {} {\bibfield  {journal} {\bibinfo  {journal} {Acta
  neuropathologica}\ }\textbf {\bibinfo {volume} {137}},\ \bibinfo {pages}
  {151--165} (\bibinfo {year} {2019})}\BibitemShut {NoStop}%
\bibitem [{\citenamefont {Helakari}\ \emph {et~al.}(2022)\citenamefont
  {Helakari}, \citenamefont {Korhonen}, \citenamefont {Holst}, \citenamefont
  {Piispala}, \citenamefont {Kallio}, \citenamefont {V{\"a}yrynen},
  \citenamefont {Huotari}, \citenamefont {Raitamaa}, \citenamefont {Tuunanen},
  \citenamefont {Kananen} \emph {et~al.}}]{helakari2022human}%
  \BibitemOpen
  \bibfield  {author} {\bibinfo {author} {\bibfnamefont {H.}~\bibnamefont
  {Helakari}}, \bibinfo {author} {\bibfnamefont {V.}~\bibnamefont {Korhonen}},
  \bibinfo {author} {\bibfnamefont {S.~C.}\ \bibnamefont {Holst}}, \bibinfo
  {author} {\bibfnamefont {J.}~\bibnamefont {Piispala}}, \bibinfo {author}
  {\bibfnamefont {M.}~\bibnamefont {Kallio}}, \bibinfo {author} {\bibfnamefont
  {T.}~\bibnamefont {V{\"a}yrynen}}, \bibinfo {author} {\bibfnamefont
  {N.}~\bibnamefont {Huotari}}, \bibinfo {author} {\bibfnamefont
  {L.}~\bibnamefont {Raitamaa}}, \bibinfo {author} {\bibfnamefont
  {J.}~\bibnamefont {Tuunanen}}, \bibinfo {author} {\bibfnamefont
  {J.}~\bibnamefont {Kananen}},  \emph {et~al.},\ }\bibfield  {title} {\enquote
  {\bibinfo {title} {Human {NREM} sleep promotes brain-wide vasomotor and
  respiratory pulsations},}\ }\href@noop {} {\bibfield  {journal} {\bibinfo
  {journal} {Journal of Neuroscience}\ }\textbf {\bibinfo {volume} {42}},\
  \bibinfo {pages} {2503--2515} (\bibinfo {year} {2022})}\BibitemShut {NoStop}%
\bibitem [{\citenamefont {Blinder}\ \emph {et~al.}(2010)\citenamefont
  {Blinder}, \citenamefont {Shih}, \citenamefont {Rafie},\ and\ \citenamefont
  {Kleinfeld}}]{blinder2010topological}%
  \BibitemOpen
  \bibfield  {author} {\bibinfo {author} {\bibfnamefont {P.}~\bibnamefont
  {Blinder}}, \bibinfo {author} {\bibfnamefont {A.~Y.}\ \bibnamefont {Shih}},
  \bibinfo {author} {\bibfnamefont {C.}~\bibnamefont {Rafie}}, \ and\ \bibinfo
  {author} {\bibfnamefont {D.}~\bibnamefont {Kleinfeld}},\ }\bibfield  {title}
  {\enquote {\bibinfo {title} {Topological basis for the robust distribution of
  blood to rodent neocortex},}\ }\href@noop {} {\bibfield  {journal} {\bibinfo
  {journal} {Proc. Natl. Acad. Sci.}\ }\textbf {\bibinfo {volume} {107}},\
  \bibinfo {pages} {12670--12675} (\bibinfo {year} {2010})}\BibitemShut
  {NoStop}%
\bibitem [{\citenamefont {Shapiro}, \citenamefont {Jaffrin},\ and\
  \citenamefont {Weinberg}(1969)}]{shapiro1969peristaltic}%
  \BibitemOpen
  \bibfield  {author} {\bibinfo {author} {\bibfnamefont {A.}~\bibnamefont
  {Shapiro}}, \bibinfo {author} {\bibfnamefont {M.}~\bibnamefont {Jaffrin}}, \
  and\ \bibinfo {author} {\bibfnamefont {S.}~\bibnamefont {Weinberg}},\
  }\bibfield  {title} {\enquote {\bibinfo {title} {Peristaltic pumping with
  long wavelengths at low {R}eynolds number},}\ }\href@noop {} {\bibfield
  {journal} {\bibinfo  {journal} {J. Fluid Mech.}\ }\textbf {\bibinfo {volume}
  {37}},\ \bibinfo {pages} {799--825} (\bibinfo {year} {1969})}\BibitemShut
  {NoStop}%
\bibitem [{\citenamefont {Akram}\ \emph {et~al.}(2023)\citenamefont {Akram},
  \citenamefont {Athar}, \citenamefont {Saeed}, \citenamefont {Razia},\ and\
  \citenamefont {Muhammad}}]{akram2023hybridized}%
  \BibitemOpen
  \bibfield  {author} {\bibinfo {author} {\bibfnamefont {S.}~\bibnamefont
  {Akram}}, \bibinfo {author} {\bibfnamefont {M.}~\bibnamefont {Athar}},
  \bibinfo {author} {\bibfnamefont {K.}~\bibnamefont {Saeed}}, \bibinfo
  {author} {\bibfnamefont {A.}~\bibnamefont {Razia}}, \ and\ \bibinfo {author}
  {\bibfnamefont {T.}~\bibnamefont {Muhammad}},\ }\bibfield  {title} {\enquote
  {\bibinfo {title} {Hybridized consequence of thermal and concentration
  convection on peristaltic transport of magneto powell--eyring nanofluids in
  inclined asymmetric channel},}\ }\href@noop {} {\bibfield  {journal}
  {\bibinfo  {journal} {Math. Methods Appl. Sci.}\ }\textbf {\bibinfo {volume}
  {46}},\ \bibinfo {pages} {11462--11478} (\bibinfo {year} {2023})}\BibitemShut
  {NoStop}%
\bibitem [{\citenamefont {Li}\ and\ \citenamefont
  {Brasseur}(1993)}]{li1993non}%
  \BibitemOpen
  \bibfield  {author} {\bibinfo {author} {\bibfnamefont {M.}~\bibnamefont
  {Li}}\ and\ \bibinfo {author} {\bibfnamefont {J.}~\bibnamefont {Brasseur}},\
  }\bibfield  {title} {\enquote {\bibinfo {title} {Non-steady peristaltic
  transport in finite-length tubes},}\ }\href@noop {} {\bibfield  {journal}
  {\bibinfo  {journal} {J. Fluid Mech.}\ }\textbf {\bibinfo {volume} {248}},\
  \bibinfo {pages} {129--151} (\bibinfo {year} {1993})}\BibitemShut {NoStop}%
\bibitem [{\citenamefont {Coenen}, \citenamefont {Zhang},\ and\ \citenamefont
  {S{\'a}nchez}(2021)}]{coenen2021lubrication}%
  \BibitemOpen
  \bibfield  {author} {\bibinfo {author} {\bibfnamefont {W.}~\bibnamefont
  {Coenen}}, \bibinfo {author} {\bibfnamefont {X.}~\bibnamefont {Zhang}}, \
  and\ \bibinfo {author} {\bibfnamefont {A.~L.}\ \bibnamefont {S{\'a}nchez}},\
  }\bibfield  {title} {\enquote {\bibinfo {title} {Lubrication analysis of
  peristaltic motion in non-axisymmetric annular tubes},}\ }\href@noop {}
  {\bibfield  {journal} {\bibinfo  {journal} {J. Fluid Mech.}\ }\textbf
  {\bibinfo {volume} {921}},\ \bibinfo {pages} {R2} (\bibinfo {year}
  {2021})}\BibitemShut {NoStop}%
\bibitem [{\citenamefont {White}(2006)}]{white2006viscous}%
  \BibitemOpen
  \bibfield  {author} {\bibinfo {author} {\bibfnamefont {F.}~\bibnamefont
  {White}},\ }\href@noop {} {\emph {\bibinfo {title} {Viscous Fluid Flow}}},\
  \bibinfo {edition} {3rd}\ ed.\ (\bibinfo  {publisher} {McGraw-Hill New
  York},\ \bibinfo {year} {2006})\BibitemShut {NoStop}%
\bibitem [{\citenamefont {Harris}\ \emph {et~al.}(2020)\citenamefont {Harris},
  \citenamefont {Millman}, \citenamefont {van~der Walt}, \citenamefont
  {Gommers}, \citenamefont {Virtanen}, \citenamefont {Cournapeau},
  \citenamefont {Wieser}, \citenamefont {Taylor}, \citenamefont {Berg},
  \citenamefont {Smith}, \citenamefont {Kern}, \citenamefont {Picus},
  \citenamefont {Hoyer}, \citenamefont {van Kerkwijk}, \citenamefont {Brett},
  \citenamefont {Haldane}, \citenamefont {del R{\'{i}}o}, \citenamefont
  {Wiebe}, \citenamefont {Peterson}, \citenamefont {G{\'{e}}rard-Marchant},
  \citenamefont {Sheppard}, \citenamefont {Reddy}, \citenamefont {Weckesser},
  \citenamefont {Abbasi}, \citenamefont {Gohlke},\ and\ \citenamefont
  {Oliphant}}]{harris2020array}%
  \BibitemOpen
  \bibfield  {author} {\bibinfo {author} {\bibfnamefont {C.~R.}\ \bibnamefont
  {Harris}}, \bibinfo {author} {\bibfnamefont {K.~J.}\ \bibnamefont {Millman}},
  \bibinfo {author} {\bibfnamefont {S.~J.}\ \bibnamefont {van~der Walt}},
  \bibinfo {author} {\bibfnamefont {R.}~\bibnamefont {Gommers}}, \bibinfo
  {author} {\bibfnamefont {P.}~\bibnamefont {Virtanen}}, \bibinfo {author}
  {\bibfnamefont {D.}~\bibnamefont {Cournapeau}}, \bibinfo {author}
  {\bibfnamefont {E.}~\bibnamefont {Wieser}}, \bibinfo {author} {\bibfnamefont
  {J.}~\bibnamefont {Taylor}}, \bibinfo {author} {\bibfnamefont
  {S.}~\bibnamefont {Berg}}, \bibinfo {author} {\bibfnamefont {N.~J.}\
  \bibnamefont {Smith}}, \bibinfo {author} {\bibfnamefont {R.}~\bibnamefont
  {Kern}}, \bibinfo {author} {\bibfnamefont {M.}~\bibnamefont {Picus}},
  \bibinfo {author} {\bibfnamefont {S.}~\bibnamefont {Hoyer}}, \bibinfo
  {author} {\bibfnamefont {M.~H.}\ \bibnamefont {van Kerkwijk}}, \bibinfo
  {author} {\bibfnamefont {M.}~\bibnamefont {Brett}}, \bibinfo {author}
  {\bibfnamefont {A.}~\bibnamefont {Haldane}}, \bibinfo {author} {\bibfnamefont
  {J.~F.}\ \bibnamefont {del R{\'{i}}o}}, \bibinfo {author} {\bibfnamefont
  {M.}~\bibnamefont {Wiebe}}, \bibinfo {author} {\bibfnamefont
  {P.}~\bibnamefont {Peterson}}, \bibinfo {author} {\bibfnamefont
  {P.}~\bibnamefont {G{\'{e}}rard-Marchant}}, \bibinfo {author} {\bibfnamefont
  {K.}~\bibnamefont {Sheppard}}, \bibinfo {author} {\bibfnamefont
  {T.}~\bibnamefont {Reddy}}, \bibinfo {author} {\bibfnamefont
  {W.}~\bibnamefont {Weckesser}}, \bibinfo {author} {\bibfnamefont
  {H.}~\bibnamefont {Abbasi}}, \bibinfo {author} {\bibfnamefont
  {C.}~\bibnamefont {Gohlke}}, \ and\ \bibinfo {author} {\bibfnamefont {T.~E.}\
  \bibnamefont {Oliphant}},\ }\bibfield  {title} {\enquote {\bibinfo {title}
  {Array programming with {NumPy}},}\ }\href {\doibase
  10.1038/s41586-020-2649-2} {\bibfield  {journal} {\bibinfo  {journal}
  {Nature}\ }\textbf {\bibinfo {volume} {585}},\ \bibinfo {pages} {357--362}
  (\bibinfo {year} {2020})}\BibitemShut {NoStop}%
\bibitem [{\citenamefont {Gjerde}(2023)}]{gjerde2023graphnics}%
  \BibitemOpen
  \bibfield  {author} {\bibinfo {author} {\bibfnamefont {I.~G.}\ \bibnamefont
  {Gjerde}},\ }\bibfield  {title} {\enquote {\bibinfo {title} {Graphnics:
  Combining {FEniCS} and {NetworkX} to simulate flow in complex networks},}\
  }\href@noop {} {\bibfield  {journal} {\bibinfo  {journal} {arXiv preprint
  arXiv:2212.02916}\ } (\bibinfo {year} {2023})}\BibitemShut {NoStop}%
\bibitem [{\citenamefont {Daversin-Catty}, \citenamefont {Gjerde},\ and\
  \citenamefont {Rognes}(2022)}]{daversin2022geometrically}%
  \BibitemOpen
  \bibfield  {author} {\bibinfo {author} {\bibfnamefont {C.}~\bibnamefont
  {Daversin-Catty}}, \bibinfo {author} {\bibfnamefont {I.~G.}\ \bibnamefont
  {Gjerde}}, \ and\ \bibinfo {author} {\bibfnamefont {M.~E.}\ \bibnamefont
  {Rognes}},\ }\bibfield  {title} {\enquote {\bibinfo {title} {Geometrically
  reduced modelling of pulsatile flow in perivascular networks},}\ }\href@noop
  {} {\bibfield  {journal} {\bibinfo  {journal} {Frontiers in Physics}\ ,\
  \bibinfo {pages} {360}} (\bibinfo {year} {2022})}\BibitemShut {NoStop}%
\bibitem [{\citenamefont {Jung}, \citenamefont {Lee},\ and\ \citenamefont
  {Kang}(2021)}]{jung2021novel}%
  \BibitemOpen
  \bibfield  {author} {\bibinfo {author} {\bibfnamefont {J.-Y.}\ \bibnamefont
  {Jung}}, \bibinfo {author} {\bibfnamefont {Y.-B.}\ \bibnamefont {Lee}}, \
  and\ \bibinfo {author} {\bibfnamefont {C.-K.}\ \bibnamefont {Kang}},\
  }\bibfield  {title} {\enquote {\bibinfo {title} {Novel technique to measure
  pulse wave velocity in brain vessels using a fast simultaneous multi-slice
  excitation magnetic resonance sequence},}\ }\href@noop {} {\bibfield
  {journal} {\bibinfo  {journal} {Sensors}\ }\textbf {\bibinfo {volume} {21}},\
  \bibinfo {pages} {6352} (\bibinfo {year} {2021})}\BibitemShut {NoStop}%
\bibitem [{\citenamefont {Seppey}\ \emph {et~al.}(2010)\citenamefont {Seppey},
  \citenamefont {Sauser}, \citenamefont {Koenigsberger}, \citenamefont
  {B{\'e}ny},\ and\ \citenamefont {Meister}}]{seppey2010intercellular}%
  \BibitemOpen
  \bibfield  {author} {\bibinfo {author} {\bibfnamefont {D.}~\bibnamefont
  {Seppey}}, \bibinfo {author} {\bibfnamefont {R.}~\bibnamefont {Sauser}},
  \bibinfo {author} {\bibfnamefont {M.}~\bibnamefont {Koenigsberger}}, \bibinfo
  {author} {\bibfnamefont {J.-L.}\ \bibnamefont {B{\'e}ny}}, \ and\ \bibinfo
  {author} {\bibfnamefont {J.-J.}\ \bibnamefont {Meister}},\ }\bibfield
  {title} {\enquote {\bibinfo {title} {Intercellular calcium waves are
  associated with the propagation of vasomotion along arterial strips},}\
  }\href@noop {} {\bibfield  {journal} {\bibinfo  {journal} {American Journal
  of Physiology-Heart and Circulatory Physiology}\ }\textbf {\bibinfo {volume}
  {298}},\ \bibinfo {pages} {H488--H496} (\bibinfo {year} {2010})}\BibitemShut
  {NoStop}%
\end{thebibliography}%

\end{document}